%% file: sensitivity_coefficients.tex
\documentclass[journal]{IEEEtran}
\usepackage{graphicx}
\usepackage[latin1]{inputenc}
\usepackage{amsmath}
\usepackage{amsfonts,dsfont,ulem}
\usepackage{amssymb}
\usepackage{algorithm}
\usepackage{algorithmic}
\usepackage{bm}
\usepackage{subfigure}
\usepackage{epstopdf}
\usepackage[french, english]{babel}
\usepackage{multirow}
\usepackage[latin1]{inputenc}
\usepackage{multicol}
\newcommand{\ubar}[1]{\text{\b{\text{$#1$}}}}   
\newtheorem{theorem}{Theorem}

\ifCLASSINFOpdf
\else
\fi
\begin{document}


\title{Efficient Computation of Sensitivity Coefficients of Node Voltages and Line Currents in Unbalanced Radial Electrical Distribution Networks}

\author{Konstantina Christakou, \textit{Member, IEEE}, Jean-Yves Le Boudec, \textit{Fellow, IEEE}, Mario Paolone, \textit{Senior Member, IEEE}, Dan-Cristian Tomozei, \textit{Member, IEEE} \thanks{Konstantina Christakou, Mario Paolone, Jean-Yves Le Boudec and Dan-Cristian Tomozei (email: konstantina.christakou@epfl.ch, jean-yves.leboudec @epfl.ch, mario.paolone@epfl.ch, dan-cristian.tomozei @epfl.ch) are with the \'{E}cole Polytechnique F\'{e}d\'{e}rale de Lausanne,
CH-1015 Lausanne, Switzerland.}
        }
        
\twocolumn[
  \begin{@twocolumnfalse}
\copyright2012 IEEE. Personal use of this material is permitted. Permission from IEEE must be obtained for all other uses, in any current or future media, including reprinting/republishing this material for advertising or promotional purposes, creating new collective works, for resale or redistribution to servers or lists, or reuse of any copyrighted component of this work in other works.
  \end{@twocolumnfalse}
  ]    
\newpage        
\maketitle

\IEEEpeerreviewmaketitle

\input{01_abstract.tex}

\input{02_introduction.tex}

\input{03_problem_formulation.tex}

\input{04_numerical_validation.tex}

\input{05_application.tex}

\input{06_conclusion.tex}

\bibliographystyle{IEEEtran}
\bibliography{mybibfile}

\end{document}

%% file: 01_abstract.tex
\selectlanguage{english}
\begin{abstract}
The problem of optimal control of power distribution systems is becoming increasingly compelling due to the progressive penetration of distributed energy resources in this specific layer of the electrical infrastructure. Distribution systems are, indeed, experiencing significant changes in terms of operation philosophies that are often based on optimal control strategies relying on the computation of linearized dependencies between controlled (e.g. voltages, frequency in case of islanding operation) and control variables (e.g. power injections, transformers tap positions). As the implementation of these strategies in real-time controllers imposes stringent time constraints, the derivation of analytical dependency between controlled and control variables becomes a non-trivial task to be solved. With reference to optimal voltage and power flow controls, this paper aims at providing an analytical derivation of node voltage and line current flows as a function of the nodal power injections and transformers tap-changers positions. Compared to other approaches presented in the literature, the one proposed here is based on the use of the $[\mathbf{Y}]$ compound matrix of a generic multi-phase radial unbalanced network. In order to estimate the computational benefits of the proposed approach, the relevant improvements are also quantified versus traditional methods. The validation of the proposed method is carried out by using both IEEE 13 and 34 node test feeders. The paper finally shows the use of the proposed method for the problem of optimal voltage control applied to the IEEE 34 node test feeder.
\end{abstract}


%% file: 02_introduction.tex
\section{Introduction}
\IEEEPARstart{O}{} ptimal controls of power systems are often based on the solution of linear problems that link control variables to controlled quantities by means of sensitivity coefficients. Typical optimization problems refer to scheduling of generators, voltage control, losses reduction, etc. So far, these categories of problems have been commonly investigated in the domain of high voltage transmission networks. However, during the past years, the increased penetration of distributed energy resources (DERs) in power distribution systems has raised the importance of developing optimal control strategies specifically applied to the operation of these networks (e.g.~\cite{singh1998power,Jenkins_Allan_Crossley_Kirschen_Strbac_2000,James:1034798,borghetti2010short,zhou2007generation,senjyu2008optimal}). Within this context, it is worth noting that the solution of optimal problems becomes of interest only if it meets the stringent time constraints required by real-time controls and imposed by the higher dynamics of these networks compared to the transmission ones.

Typical examples of optimal controls that are not yet deployed in active distribution networks (ADNs) are voltage and power flow controls. Usually this category of problems has been addressed in the literature by means of linear-approaches applied to the dependency between voltages and power flows as a function of the power injections (e.g. \cite{borghetti2010short,zhou2007generation,conti2010voltage,khatod2006novel}).

The typical approach for the solution of this class of control problems is the use of the sensitivity coefficients through an updated Jacobian matrix derived from the load flow problem \cite{peschon1968sensitivity, shirmohammadi1988compensation,wood1996power,marconato2002electric,begovic1992control}. However, from the computational point of view, the main disadvantage of such a category of methods is that, for every change in the operation conditions of the network, an updated Jacobian matrix needs to be built on the basis of the network state and needs, then, to be inverted. This procedure involves non-trivial computation constraints for the implementation in real-time centralized or decentralized controllers.

For this reason, the authors of~\cite{zhou2008simplified} have proposed the direct computation of voltages and network losses sensitivity coefficients, based on the Gauss-Seidel formulation of the load flow problem, by making use of the $[\mathbf{Z}]$ matrix of a balanced network. Also, in~\cite{conti2010voltage} it has been proposed the use of the $[\mathbf{Z}]$ matrix along with the constant-current model for loads and generators. In~\cite{khatod2006novel} the sensitivity coefficients are proposed to be calculated starting from the network branch currents. The approach presented in~\cite{bandler1980unified,bandler1982new,bandler1980unified2,ferreira1990tellegen,gurram1999sensitivity} belongs to a class of methods typically derived from circuit theory and is based on the use of the so-called adjoint network.

In order to increase the computational efficiency of this category of approaches, and to extend it to the inherent multi-phase unbalanced configuration of distribution networks, the main contribution of this paper is to provide a straightforward analytical derivation of node voltages and line currents sensitivities as a function of the power injections and transformers tap-changers positions. To this end, we propose to use the so-called $[\mathbf{Y}]$ compound matrix, which has the advantage of being sparse.

Compared to~\cite{conti2010voltage} the approach here proposed takes into account the whole admittance matrix of the network. On the other hand the analytical derivation of sensitivities in~\cite{conti2010voltage} was based on the approximated representation of the network lines where lines shunt parameters are neglected\footnote{It is important to observe that line shunt parameters are non-negligible in case of networks characterized by the presence of coaxial cables. These types of components are typical in the context of urban distribution networks}. The method presented in~\cite{khatod2006novel} always requires a base-case load flow solution and it relies on the assumption that all generators are PV nodes (i.e. with fixed voltage magnitude). Also, it does not account for the mutual coupling between different phase conductors.

The approach that appears the more general among the above listed is the one proposed in~\cite{zhou2008simplified}. However, this method depends on a pseudo-load flow approach (i.e. it makes use of a Gauss-Seidel iterative process with a fixed number of iterations) which influences the accuracy of the computed coefficients. Furthermore, compared to~\cite{zhou2008simplified} we have been able to:
\begin{itemize}
\item{generalize the problem formulation for a generic number of slack busses;}
\item{extend the computation of sensitivities to tap-changers positions (i.e. changes of slack busses reference voltages);}
\item{provide the proof that the analytical computation of sensitivities admits a unique solution for the case of radial networks and}
\item{take into account the inherent multiphase and unbalanced nature of distribution networks.}
\end{itemize}

The structure of the paper is the following: Section \ref{sec-formulation} focuses on the problem formulation by describing, in detail, the analytical procedure at the base of the proposed method. It also includes a proof of uniqueness of the solution of the linear system used to calculate the sensitivity coefficients for the case of radial networks. The same section also provides a computational cost analysis of the proposed method versus traditional approaches. Section \ref{sec-validation} validates the proposed method using the IEEE 13 and 34 node test feeders. Section \ref{sec-control} shows an application example of sensitivity coefficients related to the optimal voltage control in unbalanced distribution networks taking advantage of the possibility of computing them for all the phases. Section \ref{sec-conclusion} provides the final remarks about possible applications of the proposed method.

%% file: 03_problem_formulation.tex
\section{Problem Formulation}
\label{sec-formulation}
\subsection{Classical Computation of Sensitivity Coefficients in Power Networks}
\label{classicalform}
In this paragraph we make reference to a balanced network composed by $K$ busses.

Traditionally, there are three proposed ways to calculate the sensitivity coefficients of our interest. The first method consists of estimating them by a series of load flow calculations each performed for a small variation of a single control variable (i.e. nodal power injections, $P_l,Q_l$) \cite{borghetti2010short}:\footnote{In the rest of the paper complex numbers are denoted with a bar above (e.g. $\bar{E}$) and complex conjugates with a bar below (e.g. \ubar{E}).}
\begin{align}
\label{eq:traditional method}
\left.
\frac{\partial |\bar{E}_i|}{\partial P_l}=\frac{\Delta |\bar{E}_i|}{ \Delta P_l}
\right|_{\substack{
    \Delta P_{i, i \neq l} =0 \\
    \Delta Q_{i,i \neq l}=0}
   }\quad
\left.
\frac{\partial |\bar{I}_{ij}|}{\partial P_l}=\frac{\Delta |\bar{I}_{ij}|}{ \Delta P_l}
\right|_{\substack{
    \Delta P_{i, i \neq l} =0 \\
    \Delta Q_{i,i \neq l}=0}
   }\\ \nonumber
\left.
\frac{\partial |\bar{E}_i|}{\partial Q_l}=\frac{\Delta |\bar{E}_i|}{ \Delta Q_l} \right|_{\substack{
    \Delta P_{i, i \neq l} =0 \\
    \Delta Q_{i,i \neq l}=0}
   }
\left.\quad
\frac{\partial |\bar{I}_{ij}|}{\partial Q_l}=\frac{\Delta |\bar{I}_{ij}|}{ \Delta Q_l} \right|_{\substack{
    \Delta P_{i, i \neq l} =0 \\
    \Delta Q_{i,i \neq l}=0}
   }
\end{align}
where $\bar{E}_i$ is the direct sequence phase-to-ground voltage of node $i$ and $\bar{I}_{ij}$ is the direct sequence current flow between nodes $i$ and $j$ ($i , j \in \{1 \cdots K \}$).

The second method uses the Newton Raphson formulation of the load flow calculation to directly infer the voltage sensitivity coefficients as submatrices of the inverted Jacobian matrix (e.g. \cite{peschon1968sensitivity,shirmohammadi1988compensation,wood1996power,marconato2002electric,begovic1992control}):
\begin{equation}
J= \left[
\begin{array}{c c}
\dfrac{\partial \mathbf{P}}{\partial |\mathbf{\bar{E}}|} & \dfrac{\partial \mathbf{P}}{\partial \bm{\theta}}\\
 & \\
\dfrac{\partial \mathbf{Q}}{\partial |\mathbf{\bar{E}}|} & \dfrac{\partial \mathbf{Q}}{\partial \bm{\theta}}
\end{array}
\right].
\end{equation}
It is worth observing that such a method does not allow to compute the sensitivities against the transformers tap-changers positions. Additionally, as known, the submatrix $\dfrac{\partial \mathbf{Q}}{\partial |\mathbf{\bar{E}}|}$ is usually adopted to express voltage variations as a function of reactive power injections when the ratio of longitudinal line resistance versus reactance is negligible. It is worth noting that such an assumption is no longer applicable to distribution systems that require in addition to take into account active power injections.

A third method is derived from circuit theory. In this method Tellegen's theorem is applied in power networks and the computation of sensitivities relies on the concept of the so-called adjoint networks (e.g.~\cite{bandler1980unified,bandler1982new,bandler1980unified2,ferreira1990tellegen,gurram1999sensitivity}). This approach requires a base-case load flow solution in order to build a specific adjoint network that needs to be solved in order to infer the desired sensitivities.

\subsection{Analytical Derivation of Voltage and Current Sensitivity Coefficients}
\label{sensitivities}
This subsection contains the main analytical development of this paper related to the derivation of the voltage sensitivity coefficients \footnote{As shown in subsection \ref{sec:curr_coef} the current sensitivities can be straightforwardly derived from the voltage ones.}.
\subsubsection{Voltage Sensitivity Coefficients}
\label{volsensitivities}
the analysis starts with the voltage sensitivity coefficients. To this end, we derive mathematical expressions that link bus voltages to bus active and reactive power injections. For this purpose, a $K$-bus 3-phase generic electrical network is considered. The following analysis treats each phase of the network separately and, thus, it can be applied to unbalanced networks.

As known, the equations that link the voltage of each phase of the busses to the corresponding injected current are in total $M=3K$ and they are given by:
\begin{equation}
\label{eq:one}
[\mathbf{\bar{I}_{abc}}] = [\mathbf{\bar{Y}_{abc}}] \cdot [\mathbf{\bar{E}_{abc}}]
\end{equation}
where $[\mathbf{\bar{I}_{abc}}]=[\bar{I}^1_a,\bar{I}^1_b,\bar{I}^1_c...,\bar{I}^K_a,\bar{I}^K_b,\bar{I}^K_c]^T$, $[\mathbf{\bar{E}_{abc}}]=[\bar{E}^1_a,\bar{E}^1_b,\bar{E}^1_c...,\bar{E}^K_a,\bar{E}^K_b,\bar{E}^K_c]^T$. We denoted by $a$, $b$, $c$ the three network phases. The $[\mathbf{\bar{Y}_{abc}}]$ matrix is formed by using the so-called compound admittance matrix (e.g. \cite{arrillagapower}) as follows:
\[
\left[
\mathbf{\bar{Y}_{abc}}
\right]
=
\left[
\begin{matrix}
\scriptstyle \bar{Y}^{11}_{aa} & \scriptstyle\bar{Y}^{11}_{ab} & \scriptstyle\bar{Y}^{11}_{ac}& \cdots & \scriptstyle\bar{Y}^{1K}_{aa} & \scriptstyle\bar{Y}^{1K}_{ab} & \scriptstyle \bar{Y}^{1K}_{ac}\\
\scriptstyle\bar{Y}^{11}_{ba} & \scriptstyle\bar{Y}^{11}_{bb} & \scriptstyle\bar{Y}^{11}_{bc}& \cdots & \scriptstyle\bar{Y}^{1K}_{ba} & \scriptstyle\bar{Y}^{1K}_{bb} & \scriptstyle\bar{Y}^{1K}_{bc} \\
\scriptstyle\bar{Y}^{11}_{ca} & \scriptstyle\bar{Y}^{11}_{cb} & \scriptstyle\bar{Y}^{11}_{cc}& \cdots & \scriptstyle\bar{Y}^{1K}_{ca} & \scriptstyle\bar{Y}^{1K}_{cb} & \scriptstyle\bar{Y}^{1K}_{cc} \\
\vdots & \vdots & \vdots &\vdots & \vdots & \vdots &\vdots\\
\scriptstyle\bar{Y}^{K1}_{aa} & \scriptstyle\bar{Y}^{K1}_{ab} & \scriptstyle\bar{Y}^{K1}_{ac}& \cdots & \scriptstyle\bar{Y}^{KK}_{aa} & \scriptstyle\bar{Y}^{KK}_{ab} & \scriptstyle\bar{Y}^{KK}_{ac}\\
\scriptstyle\bar{Y}^{K1}_{ba} & \scriptstyle\bar{Y}^{K1}_{bb} & \scriptstyle\bar{Y}^{K1}_{bc}& \cdots & \scriptstyle\bar{Y}^{KK}_{ba} & \scriptstyle\bar{Y}^{KK}_{bb} & \scriptstyle\bar{Y}^{KK}_{bc} \\
\scriptstyle\bar{Y}^{K1}_{ca} & \scriptstyle\bar{Y}^{K1}_{cb} & \scriptstyle\bar{Y}^{K1}_{cc}& \cdots & \scriptstyle\bar{Y}^{KK}_{ca} & \scriptstyle\bar{Y}^{KK}_{cb} & \scriptstyle\bar{Y}^{KK}_{cc} \\
\end{matrix}
\right].
\]
In order to simplify the notation, in what follows we will assume the following correspondences:
$[\mathbf{\bar{I}_{abc}}]=[\bar{I}_1,...,\bar{I}_M]^T$,
$[\mathbf{\bar{E}_{abc}}]=[\bar{E}_1,...,\bar{E}_M]^T$ and
\[
\left[
\mathbf{\bar{Y}_{abc}}
\right]
=
\left[
\begin{matrix}
\bar{Y}_{11} & \cdots & \bar{Y}_{1M}\\
\vdots & \cdots &\vdots\\
\bar{Y}_{1M} & \cdots& \bar{Y}_{MM}
\end{matrix}
\right].
\]
For the rest of the analysis we will consider the network as composed by $S$ slack busses and $N$ busses with $PQ$ injections, (i.e. $\{1,2, \cdots M\}= \mathcal{S}\cup \mathcal{N}$, with $\mathcal{S}\cap \mathcal{N}=\emptyset$). The $PQ$ injections are considered constant and independent of the voltage. In this respect, we are assuming that for each separate perturbation of nodal power injections, the other loads/generators do not change their power set points. Therefore, the computation of the sensitivities inherently accounts for the whole response of the network in terms of variation of both active and reactive power flows. Such a consequence allows to compute the sensitivities in the close vicinity of the network state.

The link between power injections and bus voltages reads:
\begin{equation}
\label{eq:four}
\ubar{S}_{i}=\ubar{E}_{i}\sum \limits_{j \in {\mathcal{S} \cup \mathcal{N}}} \bar{Y}_{ij} \bar{E}_{j}  \quad, i \in \mathcal{N}.
\end{equation}
The derived system of equations (\ref{eq:four}) holds for all the phases of each bus of the network. Since the objective is to calculate the partial derivatives of the voltage magnitude over the active and reactive power injected in the other busses, we have to consider separately the slack bus of the system. As known, the assumptions for the slack bus equations are to keep its voltage constant and equal to the network rated value, by also fixing its phase equal to zero. Hence, for the three phases of the slack bus, it holds that:
\begin{equation}
\label{eq:five}
\frac{\partial \bar{E}_{i}}{\partial P_l}=0 \quad ,\forall i\in{\mathcal{S}}.
\end{equation}
At this point, by using equation (\ref{eq:four}) as a starting point one can derive closed-form  mathematical expressions to define and quantify voltage sensitivity coefficients with respect to active and reactive power variations in correspondence of the $N$ busses of the network. To derive voltage sensitivity coefficients, the partial derivatives of the voltages with respect to the active and reactive power $P_l$ and $Q_l$ of a bus $l \in \mathcal{N}$ have to be computed. The partial derivatives with respect to active power satisfy the following system of equations:
\begin{align}
\label{eq:six}
\mathds{1}_{\{i=l\}}=\frac{\partial \ubar{E}_i}{\partial P_l}\sum \limits_{j\in \mathcal{S}\cup \mathcal{N}} \bar{Y}_{ij} \bar{E}_{j} + \ubar{E}_{i} \sum \limits_{j \in \mathcal{N}} \bar{Y}_{ij} \frac{\partial \bar{E}_j}{\partial P_l}
\end{align}
where it has been taken into account that:
\begin{align}
\label{eq:seven}
&\frac{\partial \ubar{S}_i}{\partial P_l}=\frac{\partial \{P_i-jQ_i\}}{\partial P_l}=\mathds{1}_{\{i=l\}}.
\end{align}
The system of equations (\ref{eq:six}) is
not linear over complex numbers, but it is linear with respect to
$\frac{\partial \bar{E}_i}{\partial P_l}$,$\frac{\partial \ubar{E}_i}{\partial P_l}$, therefore it is linear over real numbers with respect to rectangular coordinates. As we show next, it has a unique solution and can therefore be used to compute the partial derivatives in rectangular coordinates to reduce the computational effort.

A similar system of equations holds for the sensitivity coefficients with respect to the injected reactive power $Q_l$. With the same reasoning, by taking into account that:
\begin{align}
\label{eq:nine}
&\frac{\partial \ubar{S}_i}{\partial Q_l}=\frac{\partial \{P_i-jQ_i\}}{\partial Q_l}=-j\mathds{1}_{\{i=l\}}
\end{align}we obtain that:
\begin{align}
\label{eq:ten}
-j\mathds{1}_{\{i=l\}}=\frac{\partial \ubar{E}_i}{\partial Q_l}\sum \limits_{j\in \mathcal{S}\cup \mathcal{N}} \bar{Y}_{ij} \bar{E}_{j} + \ubar{E}_{i} \sum \limits_{j \in \mathcal{N}} \bar{Y}_{ij} \frac{\partial \bar{E}_j}{\partial Q_l}.
\end{align}
By observing the above linear systems of equations (\ref{eq:six}) and (\ref{eq:ten}), we can see that the matrix that needs to be inverted in order to solve the system is fixed independently of the power of the $l$-th bus with respect to which we want to compute the partial derivatives. The only element that changes is the left hand side of the equations. 

Once $\frac{\partial \bar{E}_i}{\partial P_l}$,$\frac{\partial \ubar{E}_i}{\partial P_l}$ are obtained, the partial derivatives of the voltage magnitude can be expressed as:
\begin{align}
\dfrac{\partial |\bar{E}_i|}{\partial P_l}=
\dfrac{1}{|\bar{E}_i|}Re{(\ubar{E}_{i}\dfrac{\partial \bar{E}_{i}}{\partial P_l})}
\end {align}
and similar equations hold for derivatives with respect to reactive power injections.

\begin{theorem}
The system of equations (\ref{eq:six}), where $l$ is fixed and the unknowns are $\frac{\partial \bar{E}_i}{\partial P_l}$, $i\in \mathcal{N}$,
has a unique solution for every radial electrical network. The same holds for the system of equations (\ref{eq:ten}), where the unknowns are $\frac{\partial \bar{E}_i}{\partial Q_l}$, $i\in  \mathcal{N}$.
\label{theo-1}
\end{theorem}
\begin{IEEEproof}
Since the system is linear with respect to rectangular coordinates and there are as many unknowns as equations, the theorem is equivalent to showing that the corresponding homogeneous system of equations
has only the trivial solution. The homogeneous system can be written as:
 \begin{align}
\label{eq:eleven}
0=\ubar{\Delta}_i\sum \limits_{j\in \mathcal{S}\cup \mathcal{N}} \bar{Y}_{ij} \bar{E}_{j} + \ubar{E}_{i} \sum \limits_{j \in \mathcal{N}} \bar{Y}_{ij} \bar{\Delta}_j \;,\;\;\forall i \in \mathcal{N}
\end{align} where $\bar{\Delta}_i$ are the unknown complex numbers, defined for $i\in \mathcal{N}$. We want to show that $\bar{\Delta}_i=0$ for all $i\in \mathcal{N}$.
Let us consider two electrical networks with the same topology, i.e. same $[\mathbf{\bar{Y}_{abc}}]$ matrix, where the voltages are given. In the first network, the voltages are
\begin{equation}
\begin{array}{rcll}
\bar{E}_i' & = & \bar{E}_i &,\;\;\forall i \in \mathcal{S}\\
\bar{E}_i' & = & \bar{E}_i + \bar{\Delta}_i&,\;\;\forall i \in \mathcal{N}
\end{array}
\end{equation}
and in the second network they are
\begin{equation}
\begin{array}{rcll}
\bar{E}_i'' & = & \bar{E}_i &,\;\;\forall i \in \mathcal{S}\\
\bar{E}_i'' & = & \bar{E}_i - \bar{\Delta}_i&,\;\;\forall i \in \mathcal{N}
\end{array}
\end{equation}
Let $\ubar{S}'_i$ be the conjugate of the absorbed/injected power at the $i$th bus in the first network, and $\ubar{S}''_i$ in the second. Apply equation~(\ref{eq:four}) to bus $i \in \mathcal{N}$ in the first network:
\begin{equation*}
\begin{array}{rcl}
\ubar{S}_i' & = & \ubar{E}_i' \sum  \limits_{j \in {\mathcal{S} \cup \mathcal{N}}} \bar{Y}_{ij} \bar{E}_{j}' \\
&=&
(\ubar{E}_i +\ubar{\Delta}_i)
\left(
 \sum  \limits_{j \in {\mathcal{S} }} \bar{Y}_{ij} \bar{E}_{j}
 +
\sum  \limits_{j \in \mathcal{N}} \bar{Y}_{ij} (\bar{E}_{j}+\bar{\Delta}_j)
\right)
\\
&=&\ubar{E}_i \sum  \limits_{j \in {\mathcal{S} \cup \mathcal{N}}} \bar{Y}_{ij} \bar{E}_{j}+\ubar{\Delta}_i \sum  \limits_{j\in \mathcal{N}} \bar{Y}_{ij} \bar{\Delta}_{j}\\
&&+\;\; \ubar{\Delta}_i\sum \limits_{j\in \mathcal{S}\cup \mathcal{N}} \bar{Y}_{ij} \bar{E}_{j} + \ubar{E}_{i} \sum \limits_{j \in \mathcal{N}} \bar{Y}_{ij} \bar{\Delta}_j
\end{array}
\end{equation*}
Similarly, for the second network and for all busses $i \in \mathcal{N}$:
\begin{equation*}
\begin{array}{rcl}
\ubar{S}_i''
&=&\ubar{E}_i \sum  \limits_{j \in {\mathcal{S} \cup \mathcal{N}}} \bar{Y}_{ij} \bar{E}_{j}+\ubar{\Delta}_i \sum  \limits_{j\in \mathcal{N}} \bar{Y}_{ij} \bar{\Delta}_{j}\\
&&-\;\; \ubar{\Delta}_i\sum \limits_{j\in \mathcal{S}\cup \mathcal{N}} \bar{Y}_{ij} \bar{E}_{j} - \ubar{E}_{i} \sum \limits_{j \in \mathcal{N}} \bar{Y}_{ij} \bar{\Delta}_j
\end{array}
\end{equation*}
Subtract the last two equations and obtain
\begin{equation*}
\begin{array}{rcl}
\ubar{S}_i'-
\ubar{S}_i''
&=&\;\; 2 \left(\ubar{\Delta}_i\sum \limits_{j\in \mathcal{S}\cup \mathcal{N}} \bar{Y}_{ij} \bar{E}_{j} + \ubar{E}_{i} \sum \limits_{j \in \mathcal{N}} \bar{Y}_{ij} \bar{\Delta}_j\right)
\end{array}
\end{equation*}
By equation~(\ref{eq:eleven}), it follows that $\ubar{S}'_i=\ubar{S}''_i$ for all $i\in\mathcal{N}$. Thus the two networks have the same active and reactive powers at all non slack busses and the same voltages at all slack busses. As discussed in \cite{chiang1990existence} for radial distribution networks such an assumption means that the load flow problem always has a unique solution. Therefore, it follows that the voltage profile of these networks must be exactly the same, i.e. $\bar{E}_i-\bar{\Delta}_i=\bar{E}_i+\bar{\Delta}_i$ for all $i\in\mathcal{N}$ and thus $\bar{\Delta}_i=0$ for all $i \in \mathcal{N}$.
\end{IEEEproof}

\subsubsection{Current Sensitivity Coefficients}
\label{sec:curr_coef}
From the previous analysis, the sensitivity coefficients linking the power injections to the voltage variations are known. Thus, it is straightforward to express the branch current sensitivities with respect to the same power injections.
Assuming to represent the lines that compose the network by means of $\pi$ models, the current flow $\bar{I}_{ij}$ between nodes $i$ and $j$ can be expressed as a function of the phase-to-ground voltages of the relevant $i, j$ nodes as follows:
\begin{equation}
\label{eq:thirteen}
\bar{I}_{ij}=\bar{Y}_{ij}(\bar{E}_i-\bar{E}_j)
\end{equation}
where $\bar{Y}_{ij}$ is the generic element of $[\mathbf{\bar{Y}_{abc}}]$ matrix between node $i$ and node $j$.

Since the voltages can be expressed as a function of the power injections into the network busses, the partial derivatives of the current with respect to the active and reactive power injections in the network can be expressed as:
\begin{eqnarray}
\begin{split}
\label{eq:fourteen}
\frac{\partial{\bar{I}_{ij}}}{\partial{P_l}}=\bar{Y}_{ij}(\frac{\partial{\bar{E}_{i}}}{\partial{P_l}}-\frac{\partial{\bar{E}_{j}}}{\partial{P_l}})\\
\frac{\partial{\bar{I}_{ij}}}{\partial{Q_l}}=\bar{Y}_{ij}(\frac{\partial{\bar{E}_{i}}}{\partial{Q_l}}-\frac{\partial{\bar{E}_{j}}}{\partial{Q_l}})
\end{split}.
\end{eqnarray}
Applying the same reasoning as earlier, the branch current sensitivity coefficients with respect to an active power $P_l$ can be computed using the following expressions:
\begin{align}
\dfrac{\partial |\bar{I}_{ij}|}{\partial P_l}=
\dfrac{1}{|\bar{I}_{ij}|}Re{(\ubar{I}_{ij}\dfrac{\partial \bar{I}_{ij}}{\partial P_l})}.
\end{align}
Similar expressions can be derived for the current coefficients with respect to the reactive power in the busses as:
\begin{align}
\dfrac{\partial |\bar{I}_{ij}|}{\partial Q_l}=
\dfrac{1}{|\bar{I}_{ij}|}Re{(\ubar{I}_{ij}\dfrac{\partial \bar{I}_{ij}}{\partial Q_l})}.
\end{align}	 	
\subsection{Sensitivity Coefficients with respect to tap positions of transformers}
\label{tap_positions}
This subsection is devoted to the derivation of analytical expressions for the voltage sensitivity coefficients\footnote{Note as shown in Sec. \ref{sec:curr_coef} once the voltage sensitivities are obtained the ones of currents can be computed directly.} with respect to tap positions of a transformer. We assume that transformers tap-changers are located in correspondence of the slack busses of the network as for distribution networks these represent the connections to external transmission or sub-transmission networks. As a consequence, the voltage sensitivities as a function of the tap positions are equivalent to the voltage sensitivities as a function of the slack reference voltage. We assume that the transformers voltage variations due to tap position changes are small enough so that the partial derivatives considered in the following analysis are meaningful. Furthermore, we assume that the power injections at the network busses are constant and independent of the voltage.

With the same reasoning as in Sec. \ref{sensitivities}, the analysis starts in equation \eqref{eq:four}. We write $\bar E_\ell = |\bar E_\ell| e^{j\theta_\ell}$ for all busses $\ell$. For a bus $i \in \mathcal{N}$ the partial derivatives with respect to the voltage magnitude $|\bar E_k|$ of a slack bus $k \in \mathcal{S}$ are considered:
\begin{align}
\label{eq:fifteen}
-\ubar{E}_{i} \bar{Y}_{ik}e^{j\theta_k}=\ubar{W}_{ik}\sum \limits_{j\in \mathcal{S}\cup \mathcal{N}} \bar{Y}_{ij} \bar{E}_{j} + \ubar{E}_{i} \sum \limits_{j \in \mathcal{N}} \bar{Y}_{ij} \bar W_{jk},
\end{align}
where
\begin{align*}
  \bar W_{ik} := \frac{\partial \bar E_i}{\partial |\bar E_k|} =\left(\frac 1 {|\bar E_i|} \frac{\partial |\bar E_i|}{\partial |\bar E_k|} + j \frac {\partial \theta_i}{\partial |\bar E_k|}\right)\bar E_i, \ i\in \mathcal N.
\end{align*}
We have taken into account that:
\begin{align}
\label{eq:sixteen}
\frac \partial {\partial |\bar E_k|} \sum \limits_{j \in \mathcal{S}} \bar{Y}_{ij} \bar{E}_j=\bar{Y}_{ik}e^{j\theta_k}
\end{align}
and
\begin{align}
\label{eq:seventeen}
\frac{\partial \ubar{S}_i}{\partial |\bar{E}_k|}=0.
\end{align}
The derived system of equations (\ref{eq:fifteen}) is linear with respect to $\ubar{W}_{ik}$ and $\bar{W}_{ik}$, and has the same associated matrix as the system in \eqref{eq:six}. Since the resulting homogeneous system of equations is identical to the one in \eqref{eq:eleven}, by Theorem~\ref{theo-1} it has a unique solution.

After resolution of \eqref{eq:fifteen}, we find that the sensitivity coefficients with respect to the tap position of the transformer at bus $k$ are given by
\begin{equation}
  \label{eq:tap-sens}
  \frac {\partial |\bar E_i|}{\partial |\bar E_k|} = |\bar E_i| Re{\left(\frac {\bar W_{ik}}{\bar E_i}\right)}.
\end{equation}

\subsection{Computational Cost Analysis for Voltage Sensitivities with respect to PQ injections}
The aim of this subsection is to show the computational advantage of the proposed method compared to the classical approach with respect to the computation of  voltage sensitivities as a function of power injections only\footnote{As already pointed out in Sec.\ref{classicalform} traditional Jacobian based sensitivity computations do not account indeed the variations of tap-changers.}. Furthermore, the two methods are applied to the IEEE $13$ and $34$ node test feeders and compared in terms of CPU time necessary to calculate the voltage sensitivity coefficients.

We are assuming that:
\begin{enumerate}
\item{there are loads/injections in all three phases of the system and}
\item{the phasors of phase-to-ground voltages in all the network are known (e.g. coming from a state estimation process \cite{abur2004power}).}
\end{enumerate}

In the following table, Algorithm 1 shows the steps required to calculate the voltage sensitivity coefficients using the traditional method and Algorithm 2 shows the corresponding steps using the analytical method proposed here.

For the traditional method an updated Jacobian needs to be built, and its inverse will provide the desired voltage sensitivities. For the analytical method the corresponding steps refer to invert a square matrix of size $2N$ (as reported in Section \ref{volsensitivities} $N$ refers to the number of network busses with $PQ$ injections) and multiply the inverse matrix with one column vector for each $PQ$ bus in the network.
\begin{algorithm}
\label{alg:Jacobian}
\caption{Computation of voltage sensitivity coefficients using the Jacobian method}
\begin{algorithmic}[1]
    \STATE  build Jacobian matrix associated to the Newton Raphson method
    \STATE  invert matrix $J$ of size $2N \times 2N$
    \STATE  extract the sub-matrices corresponding to the desired sensitivity coefficients
\end{algorithmic}
\end{algorithm}
\vspace{-20pt}
\begin{algorithm}
\label{alg:Analytical}
\caption{Computation of voltage sensitivity coefficients using the analytical method}
\begin{algorithmic}[1]
    \STATE  build the matrix of the linear system of equations
    \STATE  invert matrix of size $2N \times 2N$
    \STATE  do $N$ multiplications of the inverse matrix with vectors of size $2N \times 1$
\end{algorithmic}
\end{algorithm}

In Table \ref{table:comparison} the mean CPU time necessary to calculate the voltage sensitivity coefficients is presented for the IEEE $13$ and $34$ node test feeders respectively, when 1000 iterations of the method are executed. It can be observed that the analytical approach exhibits an improvement of performance which is of $2.34$ for the IEEE $13$ node test feeder and $2.52$ for the IEEE $34$ node test feeder. In the same table the relevant $95\%$ confidence intervals are also reported for the computation of the coefficients for the two benchmark feeders. One can observe the advantage of the proposed analytical method as the number of busses in the network increases. It is worth observing that such an improvement depends not only on the number of busses but also on the network topology (i.e. sparsity of the $[\mathbf{Y}]$ admittance matrix).

\begin{table}[H]
\caption{CPU time necessary for calculating voltage sensitivity coefficients in the IEEE 13 and the 34 node test feeders when all phases of all busses have loads}
\begin{center}
\begin{tabular}{ c | c | c | c | c |}
         \cline{2-4}
        & \multicolumn{1}{ c|}{\textbf{Jacobian}}& \multicolumn{1}{ c|}{\textbf{Analytical}} & \multicolumn{1}{ c|}{\textbf{ratio}}\\\hline
        \multicolumn{1}{|c|}{\textbf{13 bus feeder} } & $28.8 \pm 0.18$ msec & $12.5 \pm 0.43$ msec & $2.34$  \\\hline
        \multicolumn{1}{|c|}{\textbf{34 bus feeder}}  & $209.8 \pm 1.30$ msec & $83.4 \pm 0.59$ msec & $2.52$ \\\hline
\end{tabular}
\end{center}
\label{table:comparison}
\end{table}
%

%% file: 04_numerical_validation.tex
\section{Numerical validation}
\label{sec-validation}
\begin{figure}[H]
\begin{center}
  \includegraphics[width=0.92\linewidth, clip=true, trim=10 15 10 70]{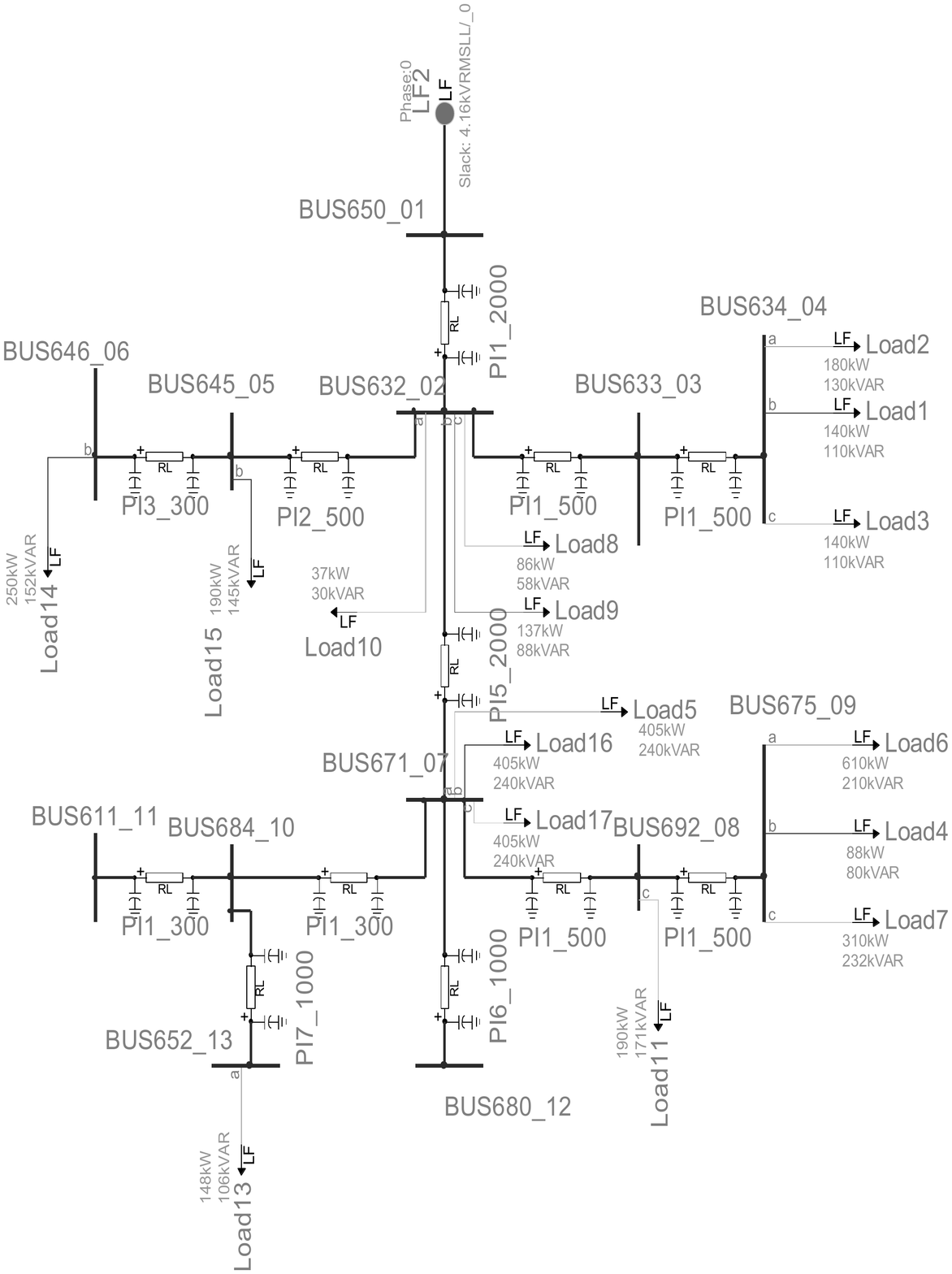}
  \caption{IEEE 13 node test feeder represented in the EMTP-RV simulation environment.}\label{fig:IEEE13}
  \end{center}
\end{figure}

The numerical validation of the proposed method for the computation of voltage/current sensitivities is performed with two different approaches. In particular, as the inverse of the load flow Jacobian matrix provides the voltage sensitivities, the comparison reported below makes reference to such a method for the voltage sensitivities only.
On the contrary, as the inverse of the load flow Jacobian matrix does not provide current sensitivity coefficients, their accuracy is evaluated by using a numerical approach where the load flow problem is solved by applying small injection perturbations into a given network (see Section \ref{classicalform}). A similar approach is deployed to validate the sensitivities with respect to tap positions of the transformers, i.e. small perturbations of the voltage magnitude of one phase of the slack bus and solution of the load flow problem.
Fig.\ref{fig:IEEE13} shows the IEEE 13 nodes test feeder implemented in the EMTP-RV simulation environment (\cite{mahseredjian1993new,mahseredjian2008simulation,mahseredjian2007new}) adopted to perform the multiphase load flow.

\begin{figure}[H]
\begin{center}
\leavevmode
\subfigure{\label{fig:3a}\includegraphics[width=1\linewidth,clip=true]{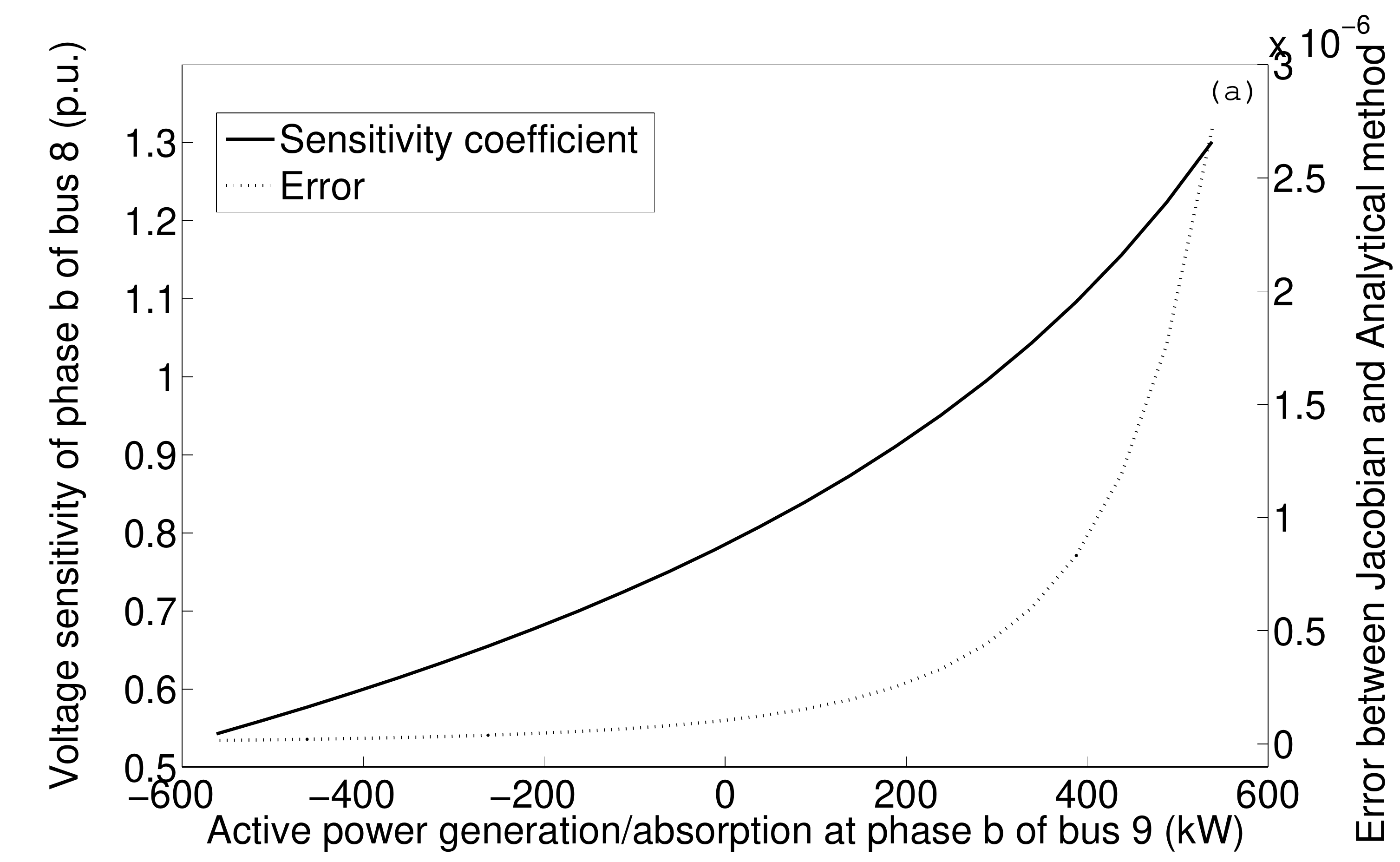}}
\vspace{4pt}
\subfigure{\label{fig:3b}\includegraphics[width=1\linewidth,clip=true,trim=0 0  0 0]{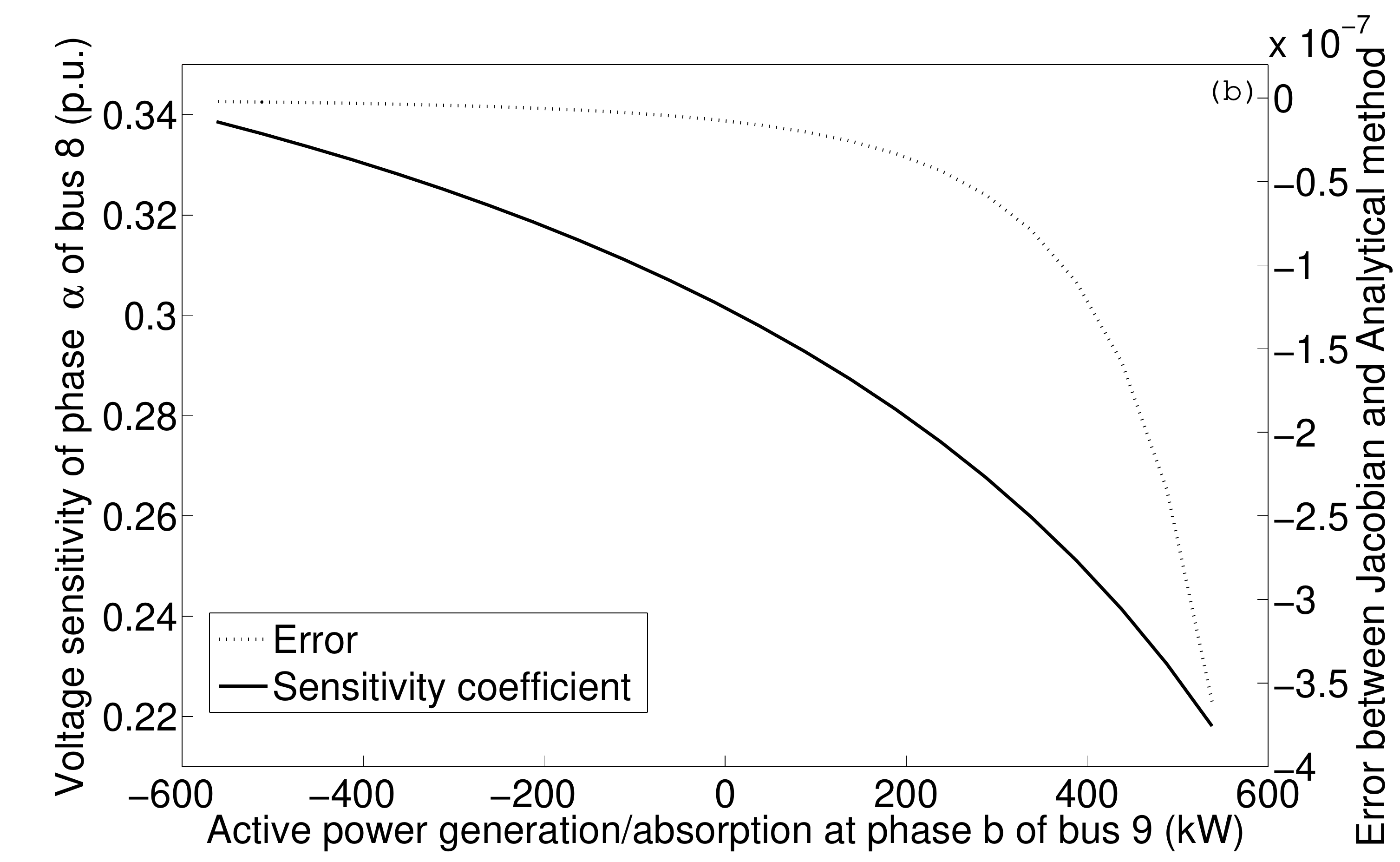}}
\vspace{4pt}
\subfigure{\label{fig:3c}\includegraphics[width=1\linewidth,clip=true]{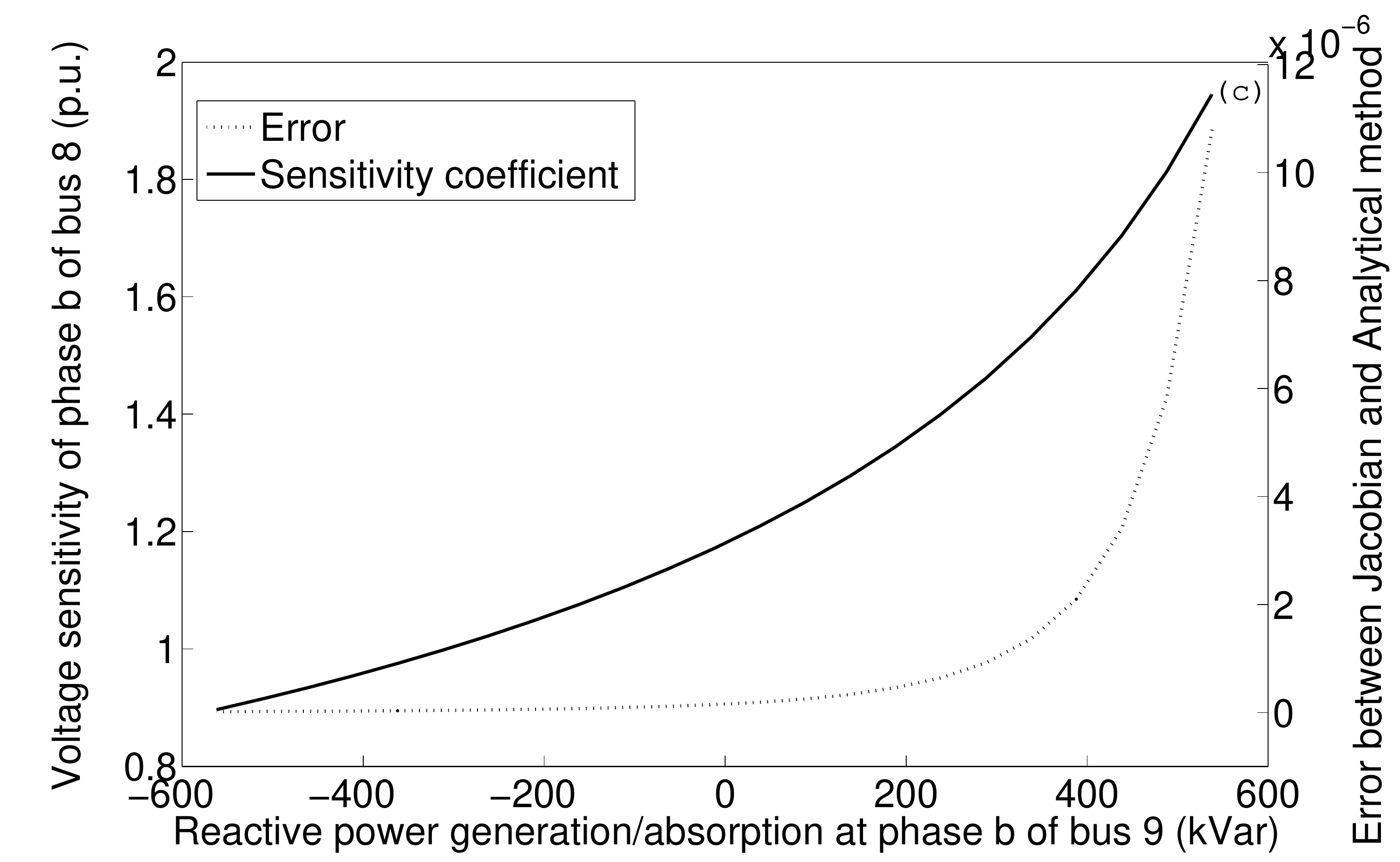}}
\vspace{4pt}
\subfigure{\label{fig:3d}\includegraphics[width=1\linewidth,clip=true]{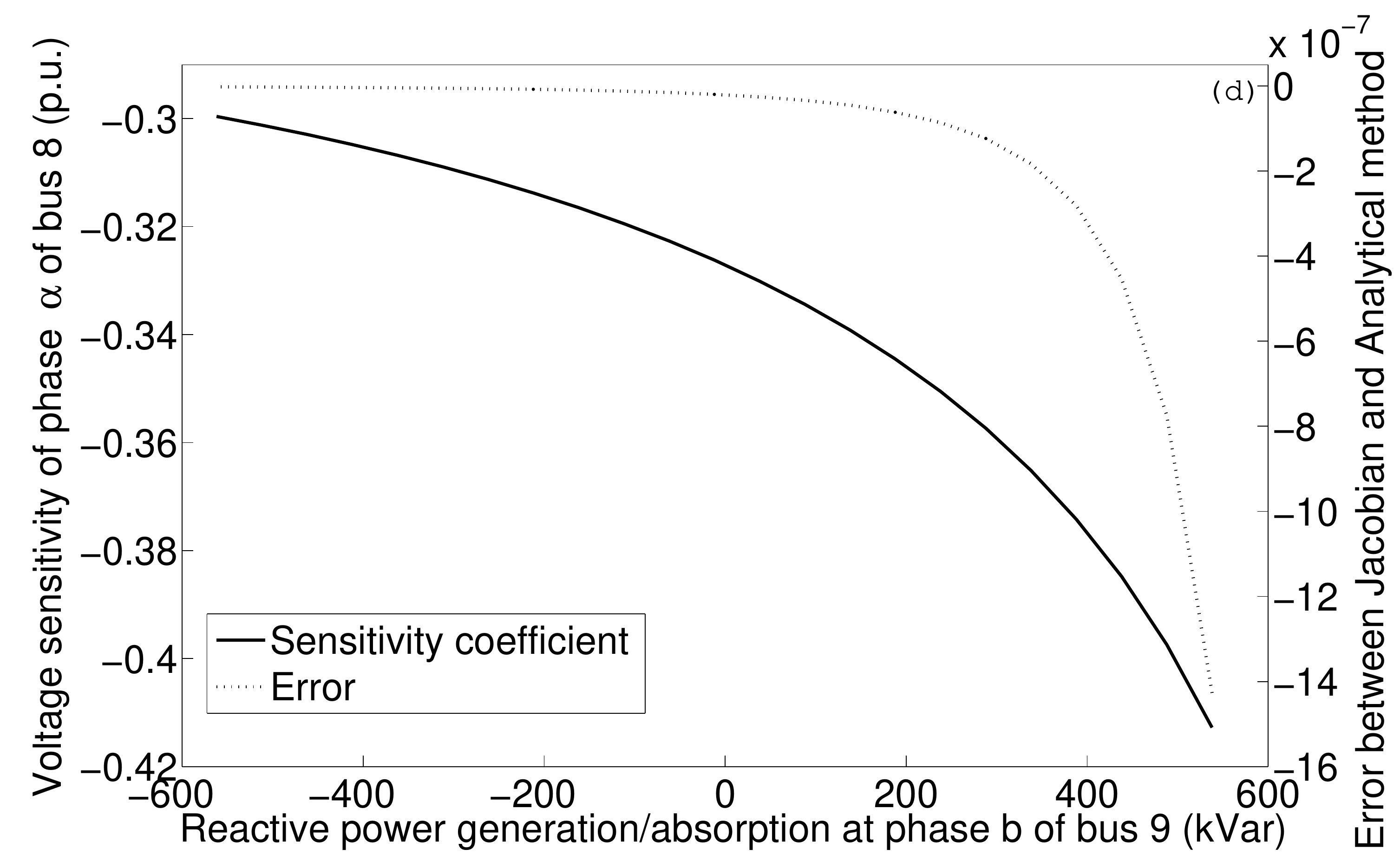}}
\end{center}
\caption{Voltage sensitivity coefficient of phase $a$ and $b$  of bus 8 with respect to active and reactive power generation/absorption at phase $b$ of bus 9.} \label{fig:8PQ9}
\end{figure}
For the sake of brevity we limit the validation of the proposed method to a reduced number of busses exhibiting the largest voltage sensitivity against $PQ$ load/injections. In particular, we refer to the variation of voltages at bus $8$ with respect to load/injection in bus $9$, i.e.
\begin{equation}
\dfrac{\partial |\bar{E}^a_8|}{\partial P^b_9}, \dfrac{\partial |\bar{E}^b_8|}{\partial P^b_9}, \dfrac{\partial |\bar{E}^a_8|}{\partial Q^b_9}, \dfrac{\partial |\bar{E}^b_8|}{\partial Q^b_9}
\nonumber
\end{equation}
In Fig.\ref{fig:3a} the voltage sensitivity of phase \textit{b} bus 8 is shown with respect to active power absorption and generation at phase \textit{b} of bus 9.
We assume the convention that positive values of P and Q denote power absorption, whereas negative values correspond to power generation. Fig.\ref{fig:3b} shows for the same busses as Fig.\ref{fig:3a}, the same sensitivity but referring to voltage and power belonging to different phases. Additionally, Fig. \ref{fig:3c} and \ref{fig:3d} show the voltage sensitivity of bus 8 with respect to reactive power absorption and generation at bus 9. In all these four figures the dashed line represents the relative error between the traditional approach (i.e. based on the inverse of the Jacobian matrix) and the analytical method proposed here. As it can be observed, the overall errors are in the order of magnitude of $10^{-6}.$
\begin{figure}[b!]
\begin{center}
\leavevmode
\subfigure{\label{fig:cur10_13p13}\includegraphics[width=1\linewidth,clip=true, trim= 0 0 0 10]{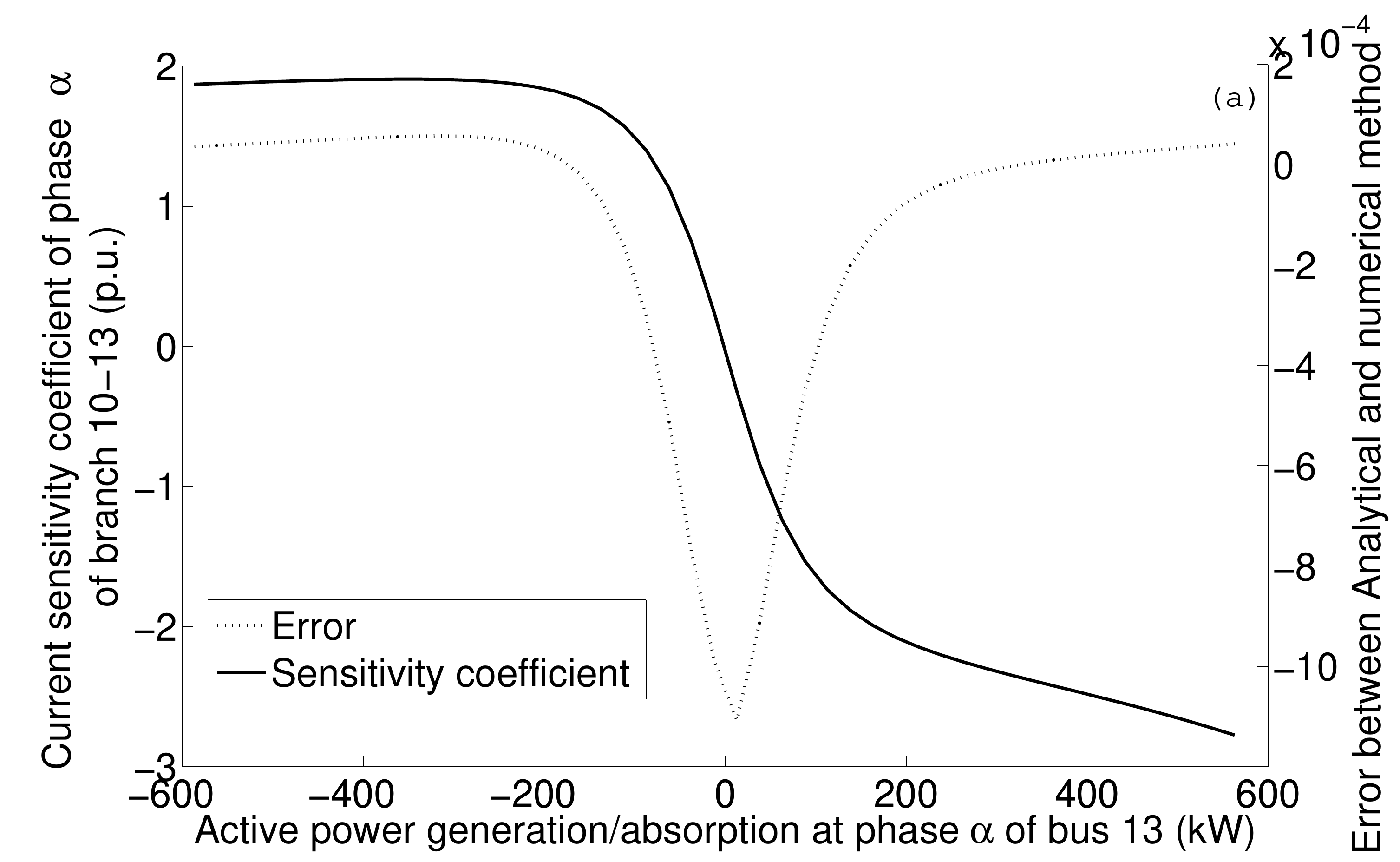}}

\subfigure{\label{fig:cur10_13q13}\includegraphics[width=1\linewidth,clip=true, trim= 0 0 0 10]{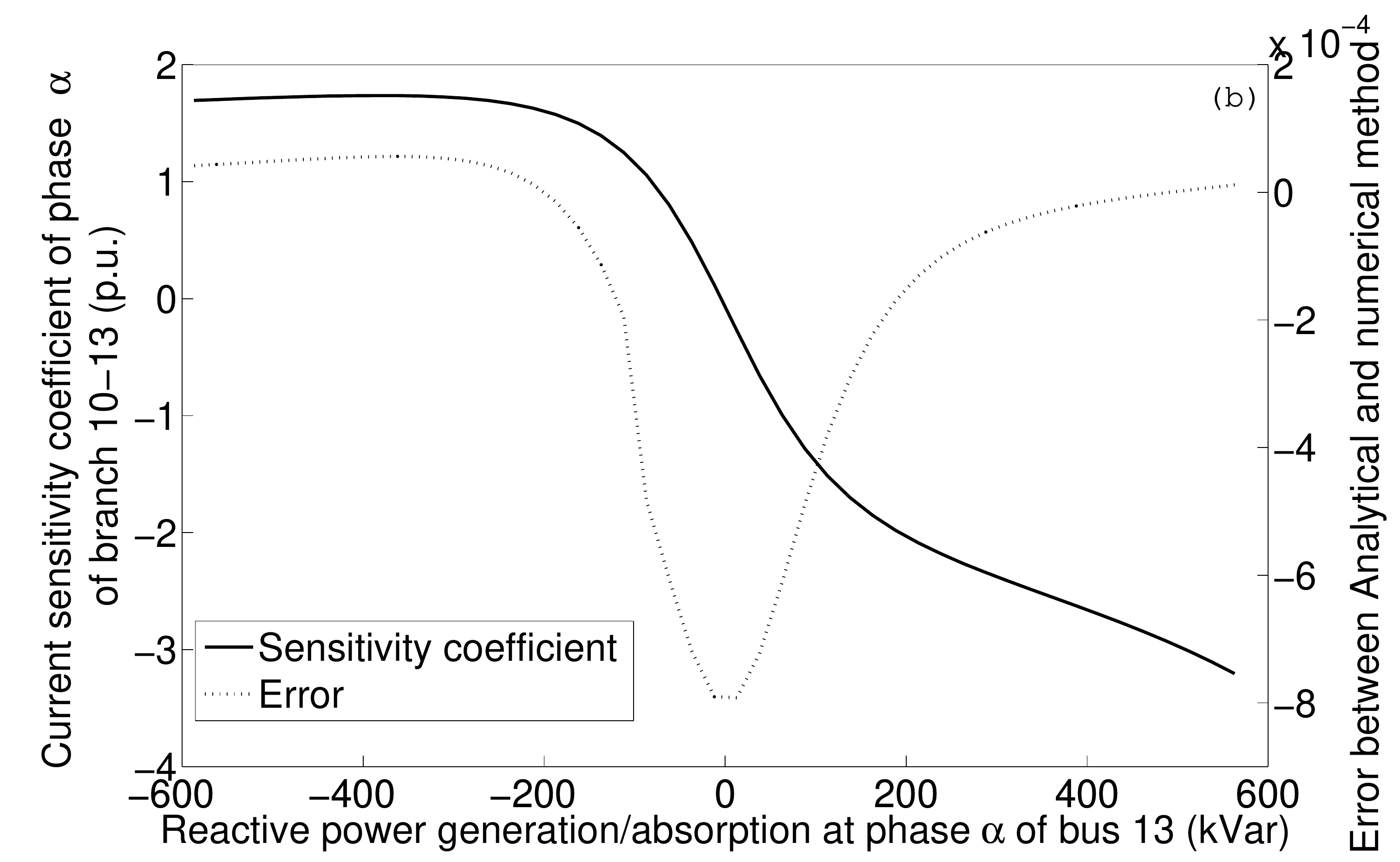}}
\end{center}
\caption{Current sensitivity coefficients of phase $a$ of branch 10-13 with respect to power generation/absorption at phase $a$ of node 13.} \label{fig:cur10_13_13}
\end{figure}
In Fig.\ref{fig:cur10_13p13} and Fig.\ref{fig:cur10_13q13} the current sensitivity coefficient of phase \textit{a} of branch $10-13$ is presented with respect to active and reactive power absorption/generation at phase \textit{a} of bus 13. In the same figures, the dashed lines represent the relative error between the analytical values and the numerical ones. Even for these coefficients extremely low errors are obtained.

\begin{figure}[h!]
\begin{center}
\leavevmode
\subfigure[Voltage sensitivity coefficient of phase $a$ of bus 7 with respect to transformer's tap position at phase $a$ of the slack bus.]{\label{fig:tapa}\includegraphics[width=0.9\linewidth,clip=true,trim= 55 100 60 100]{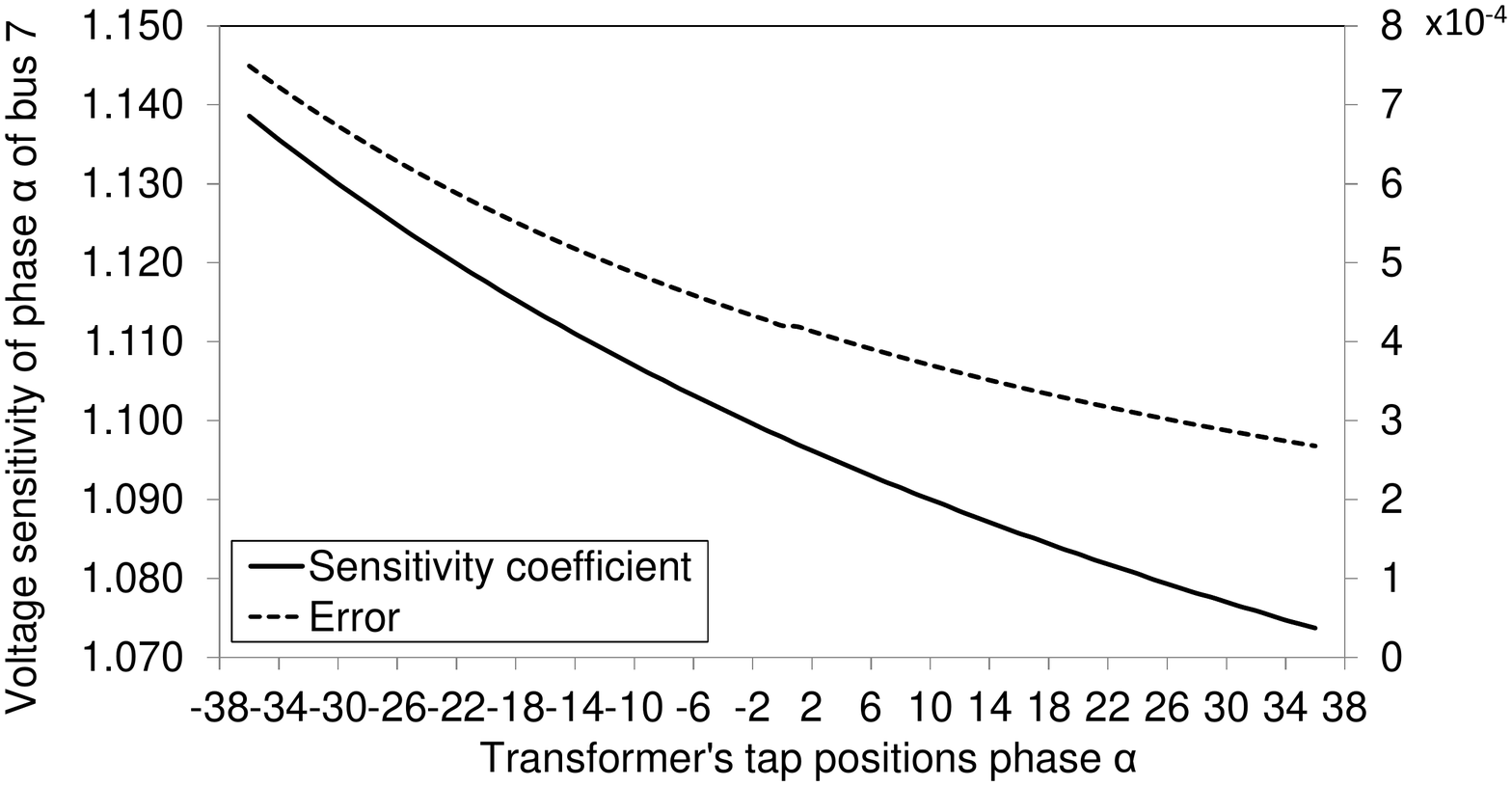}}
\subfigure[Voltage sensitivity coefficient of phase $a$ of bus 7 with respect to transformer's tap position at phase $b$ of the slack bus.]{\label{fig:tapb}\includegraphics[width=0.9\linewidth,clip=true,trim= 55 100 50 100]{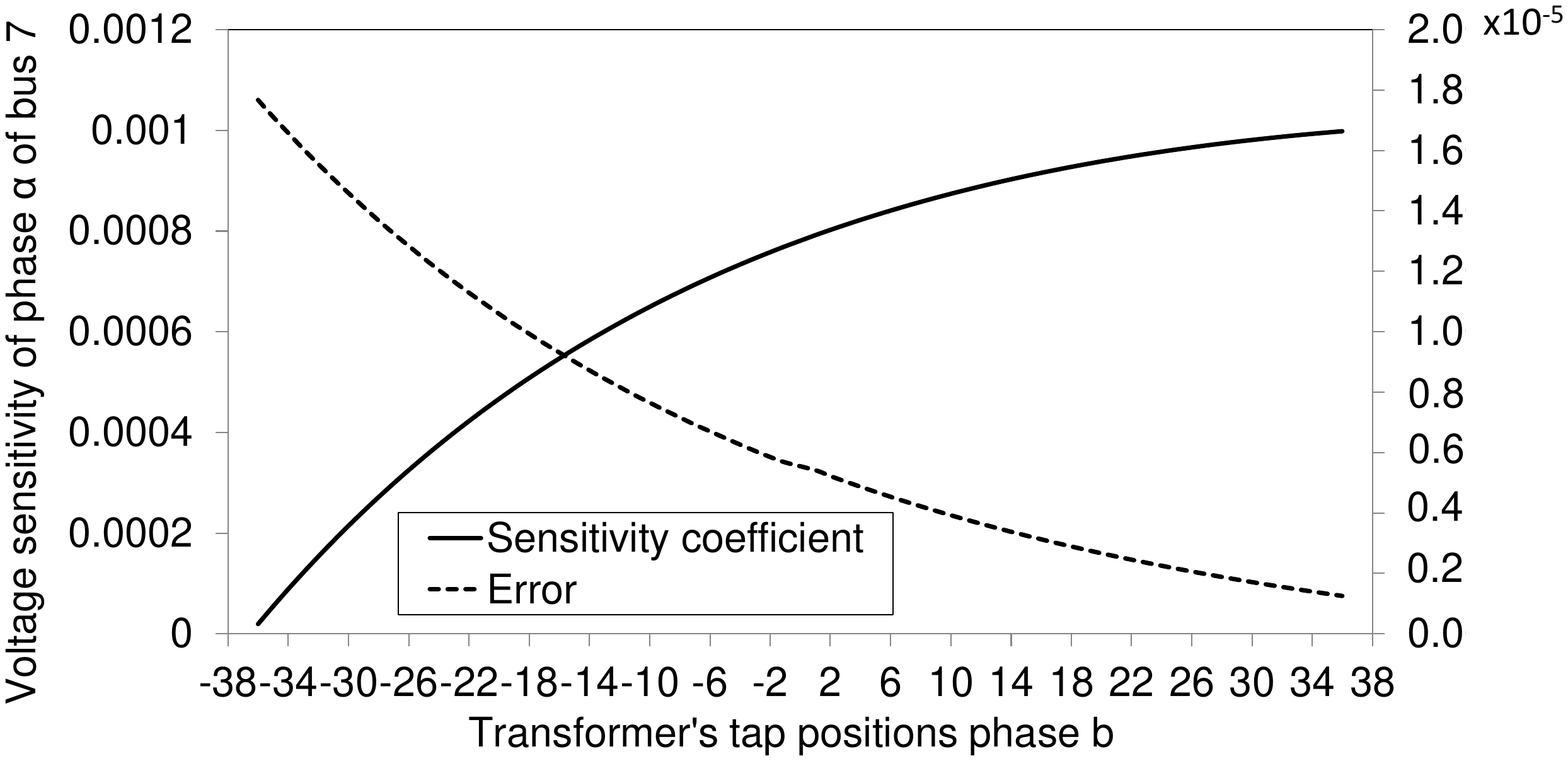}}
\subfigure[Voltage sensitivity coefficient of phase $a$ of bus 7 with respect to transformer's tap position at phase $c$ of the slack bus.]{\label{fig:tapc}\includegraphics[width=0.9\linewidth,clip=true,trim= 55 100 50 100]{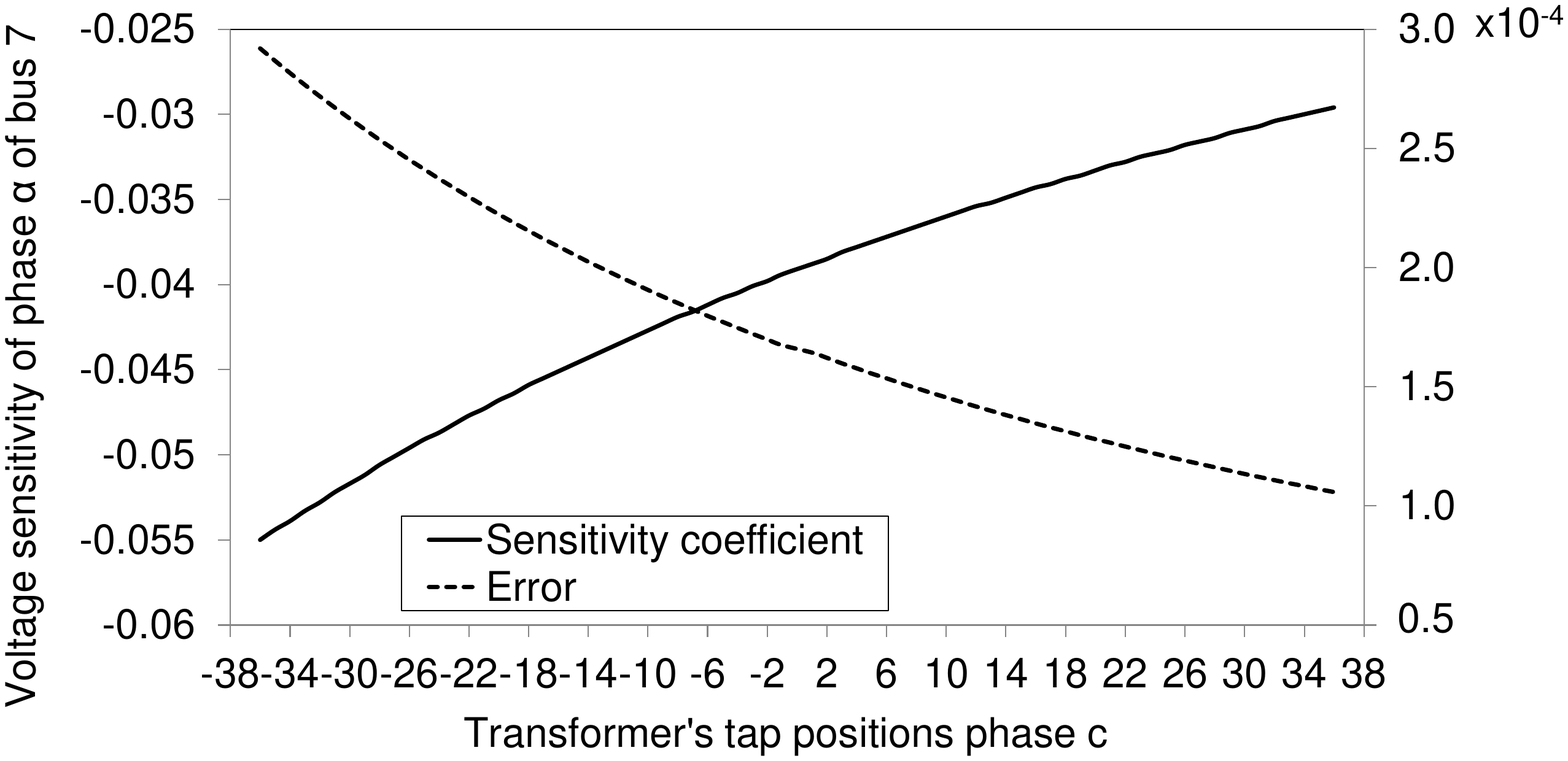}}
\end{center}
\caption{Voltage sensitivity coefficient of phase $a$ of bus 7 with respect to transformer's tap positions.} \label{fig:slack_sens}
\end{figure}

Concerning the validation of voltage sensitivities against tap-changer positions, we have made reference to the IEEE 13 node test feeder where the slack bus and therefore the primary substation transformer is placed in correspondence of node 1. We assume to vary the slack bus voltage of $\pm 6\% $ over 72 tap positions (where position "0" refers to the network rated voltage). In Fig. \ref{fig:slack_sens} the sensitivity of voltage in phase \textit{a} of bus $7$ is shown w.r.t. the tap positions in phase \textit{a}, \textit{b} and \textit{c} of the slack.
Also, in this case the difference between the analytically inferred sensitivities and the numerical computed ones is negligible (i.e. in the order of magnitude of $10^{-4}$).

\begin{figure}[H]
\begin{center}
\leavevmode
\subfigure[Voltage sensitivity coefficients $\frac{\partial |\bar{E}^a_i|}{\partial P^a_{13}}$ with respect to active power absorption at phase $a$ of node 13 as a function of the distance from the slack bus.]{\includegraphics[width=0.715\linewidth,clip=true,trim=58 250 55 250]{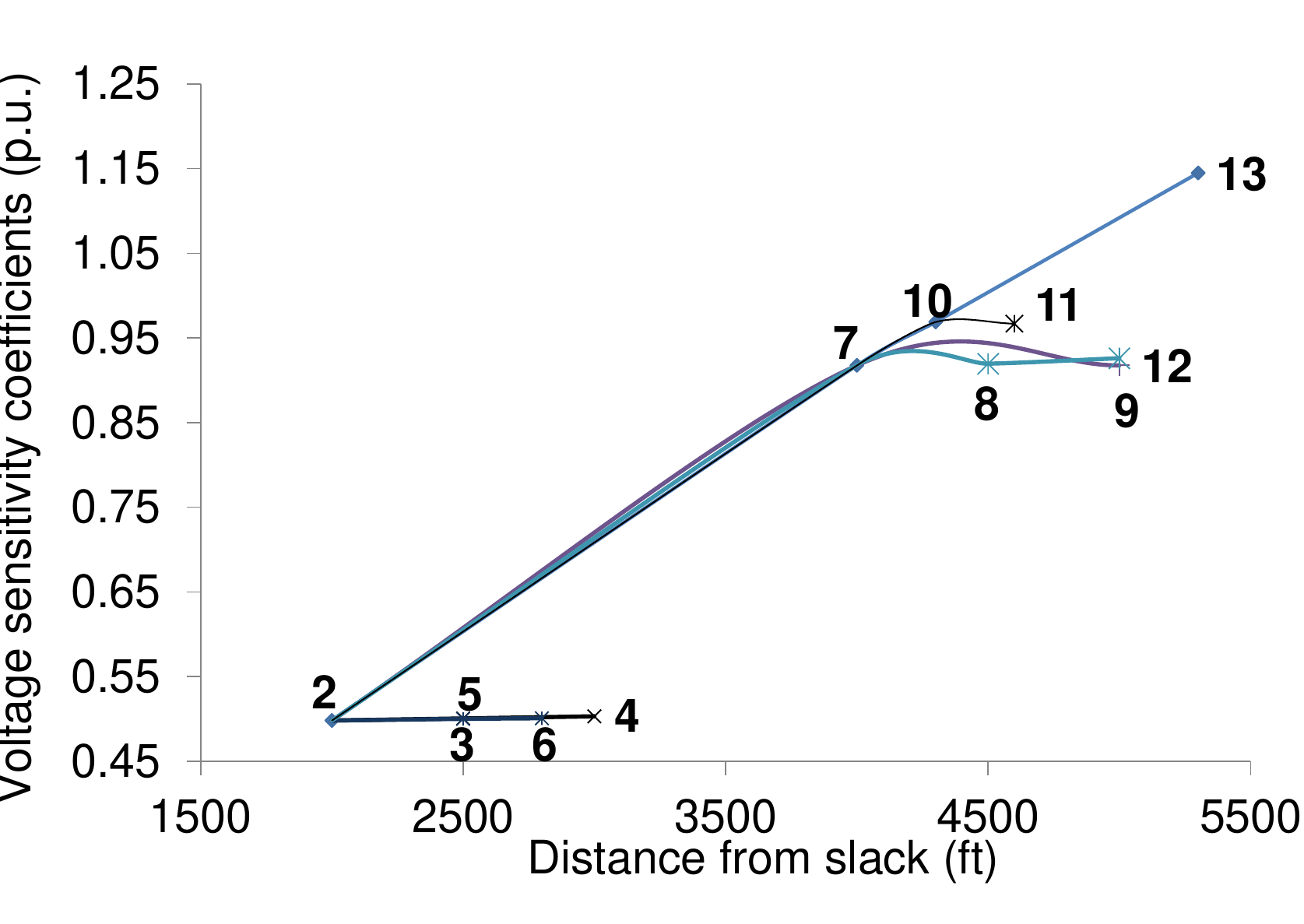}}
\subfigure[Voltage sensitivity coefficients $\frac{\partial |\bar{E}^b_i|}{\partial P^a_{13}}$ with respect to active power absorption at phase $a$ of node 13 as a function of the distance from the slack bus.]{\includegraphics[width=0.715\linewidth,clip=true,trim=60 250 55 250]{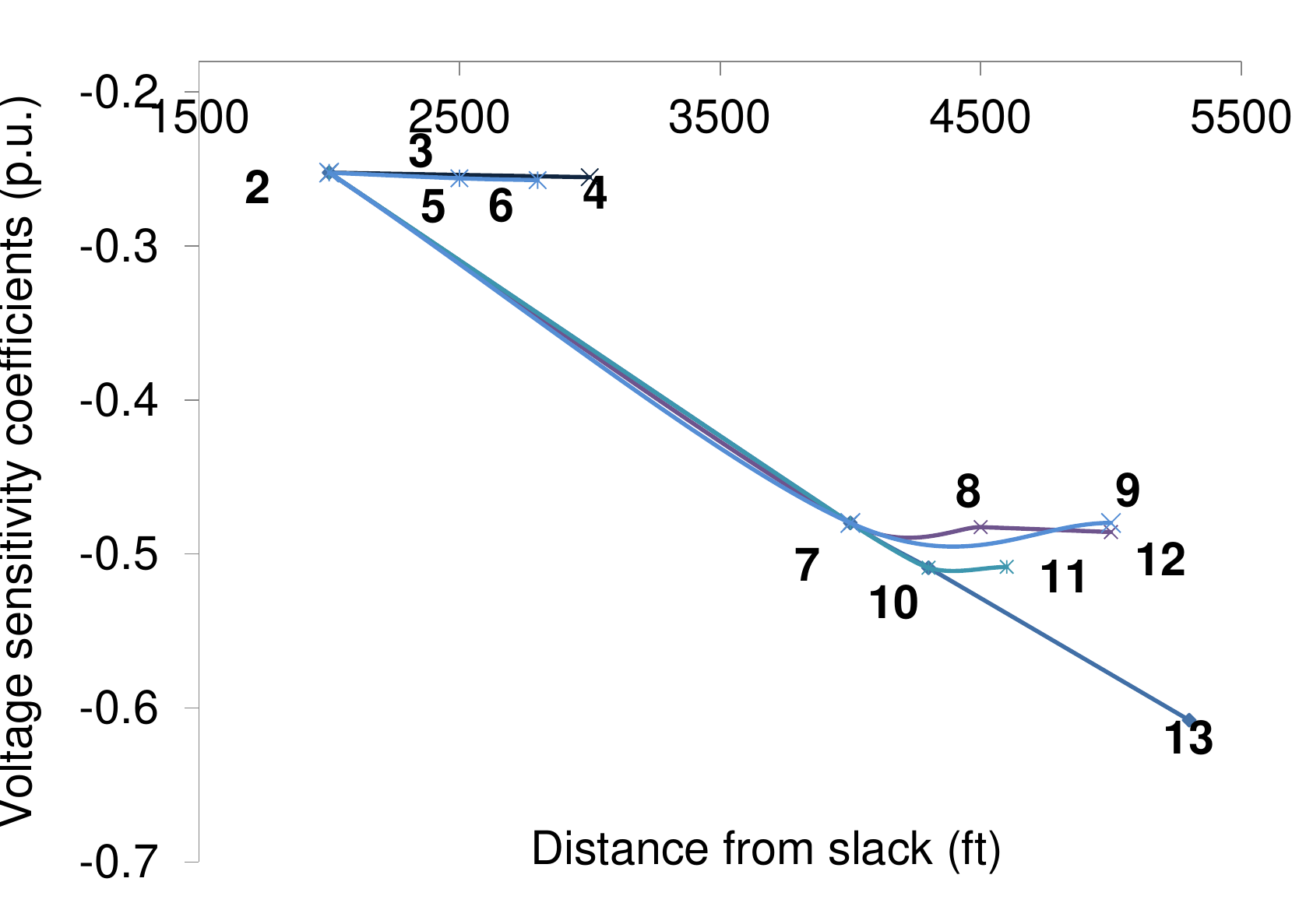}}
\subfigure[Voltage sensitivity coefficients $\frac{\partial |\bar{E}^a_i|}{\partial Q^a_{13}}$ with respect to reactive power absorption at phase $a$ of node 13 as a function of the distance from the slack bus.]{\includegraphics[width=0.715\linewidth,clip=true,trim=60 250 55 250]{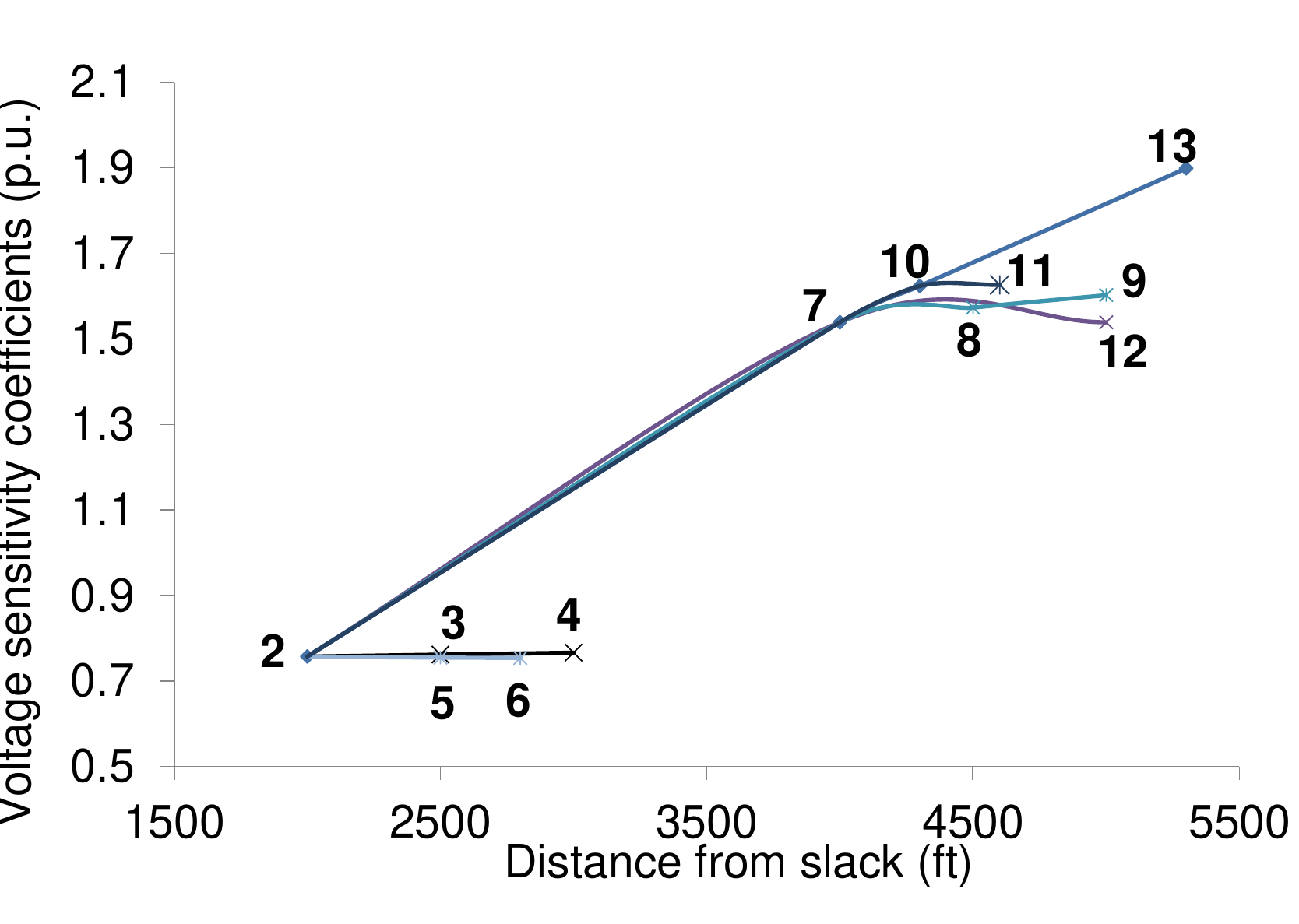}}
\subfigure[Voltage sensitivity coefficients $\frac{\partial |\bar{E}^b_i|}{\partial Q^a_{13}}$ with respect to reactive power absorption at phase $a$ of node 13 as a function of the distance from the slack bus.]{\includegraphics[width=0.715\linewidth,clip=true,trim=60 250 55 250]{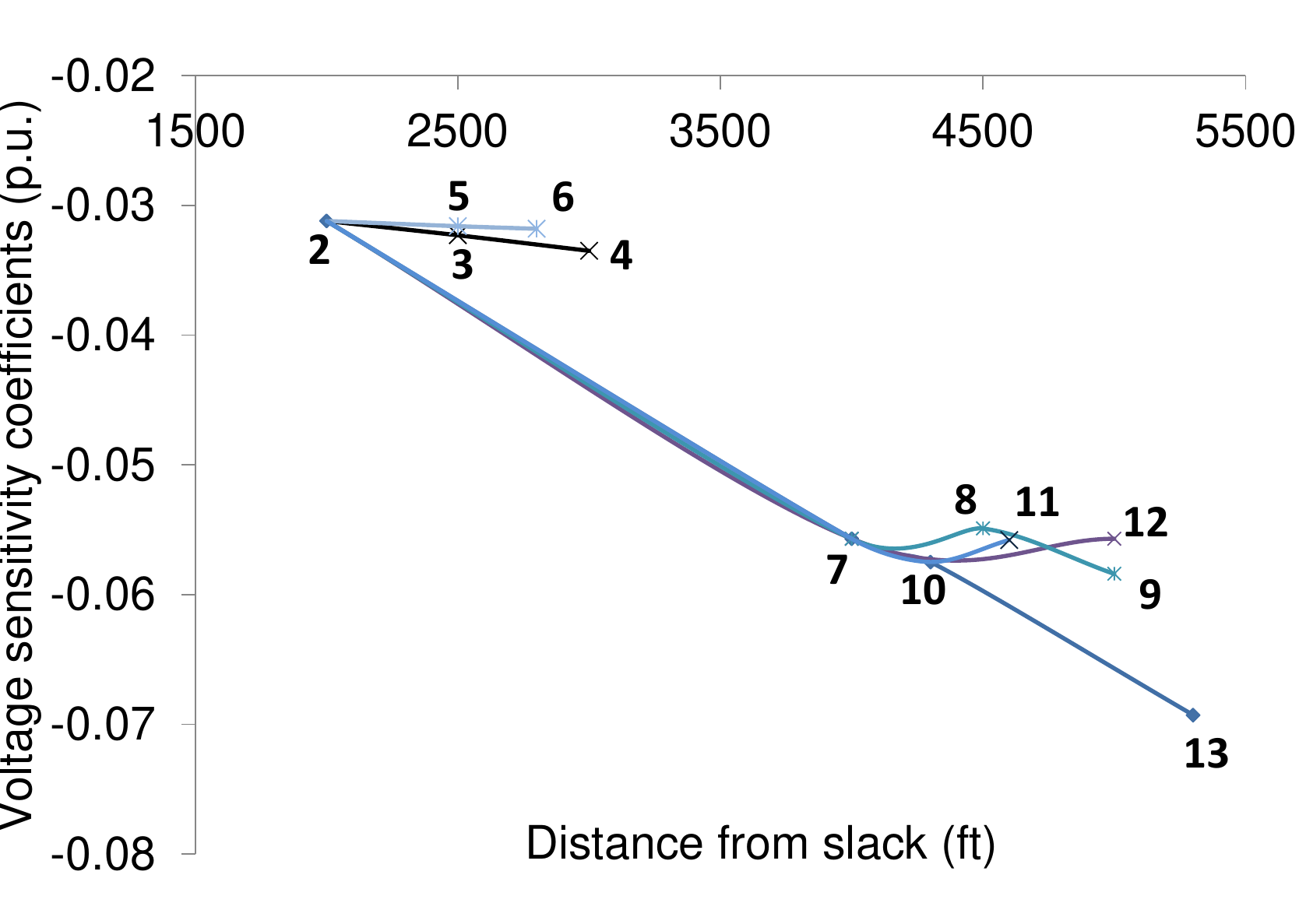}}
\end{center}
\caption{Voltage sensitivity coefficients with respect to power absorption at phase $a$ of bus 13 as a function of the distance from the slack bus.} \label{fig:voltdis}
\end{figure}

It is worth observing that for the case of the voltage sensitivities, coefficients that refer to the voltage variation as a function of a perturbation (power injection or tap-changer position) of the same phase, show the largest coupling although a non-negligible cross dependency can be observed between different phases.

Finally, Fig.\ref{fig:voltdis} depicts the variation of voltage sensitivity coefficients in all the network with respect to active and reactive power absorption at phase $a$ of bus 13 as a function of the distance from the slack bus in feet.

This type of representation allows to observe the overall network behavior against specific PQ busses absorptions/injections. In particular, we can see that larger sensitivities are observed when the distance between the considered voltage and the slack bus increases. Furthermore, a lower, but quantified dependency between coefficients related to different phases, can be observed. Also, as expected, reactive power has a larger influence on voltage variations although the active power exhibits a non negligible influence.

From the operational point of view it is worth observing that, figures as Fig.\ref{fig:voltdis}, provide to network operators an immediate view of the response of the electrical network against specific loads/injections that could also be used for closed loop control or contingency analysis.

\begin{figure}[H]
\begin{center}
  \includegraphics[width=1\linewidth, clip=true,trim= 10 10 10 10 ]{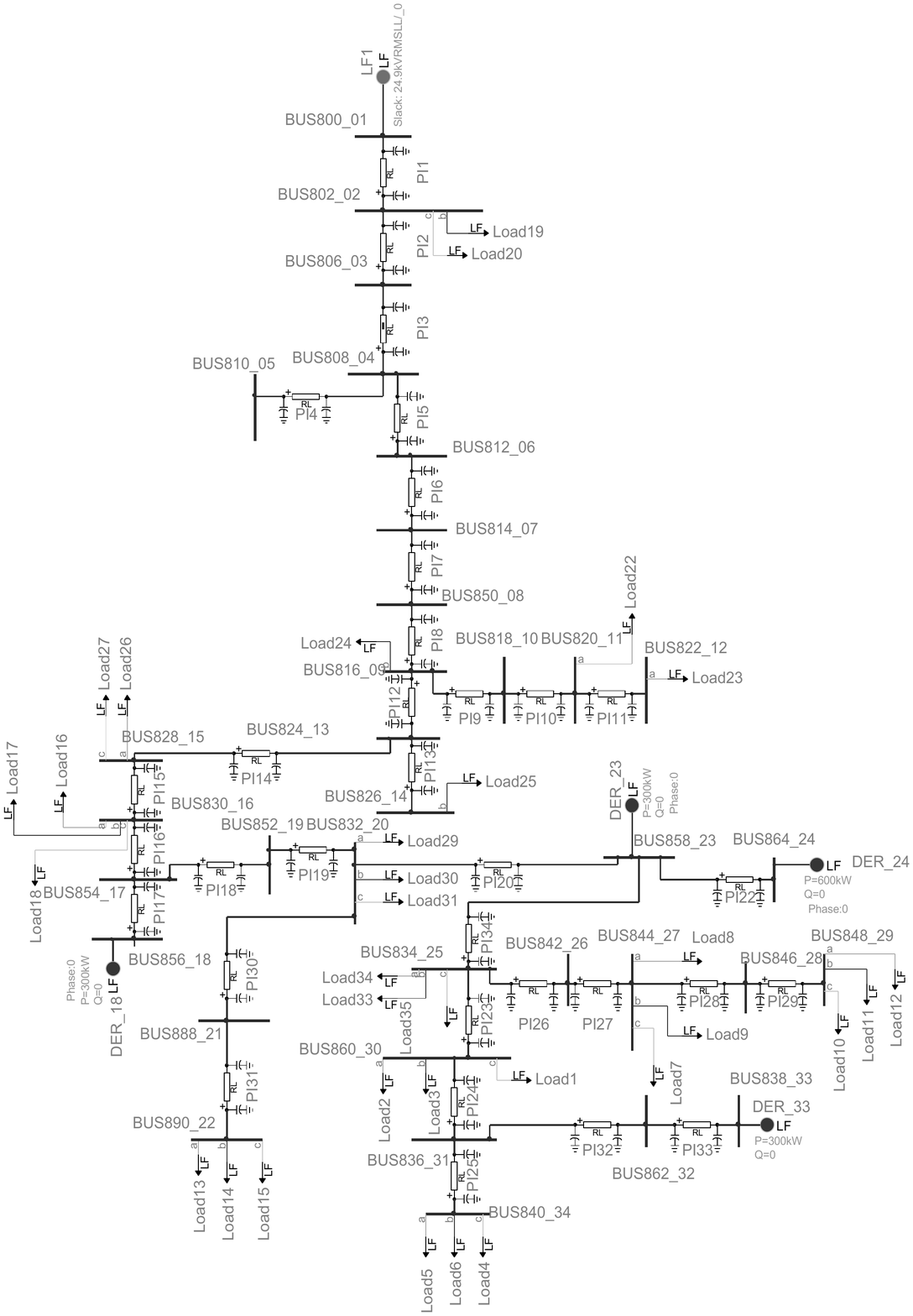}
  \caption{IEEE 34 node test feeder represented in the EMTP-RV simulation environment.}\label{fig:IEEE34}
  \end{center}
\end{figure}

%% file: 05_application.tex
\section{Application of the proposed procedure to the problem of Optimal Voltage Control}
\label{sec-control}
For the application part, the IEEE $34$ test node feeder is considered as depicted in Fig.\ref{fig:IEEE34}. In busses $18$, $23$, $24$ and $33$ we assume to have distributed energy resources that the Distribution Network Operator (DNO) can control in terms of active and reactive power. Their initial operating values, as well as their rated power outputs, are shown in Table \ref{table:initialDER}. Furthermore, the DNO has control on the transformer's tap positions.
\begin{table}[H]
\begin{center}
\caption{Initial and maximum operational set points of the DERs and the tap-changers in the 34 test node feeder}
\label{table:initialDER}
\begin{tabular}{ c | c | c | c | c | c |}
        \cline{2-6}
        & \multicolumn{1}{c|}{\textbf{$\mathbf{P_{init}(kW)}$}}&\multicolumn{1}{c|}{\textbf{$\mathbf{P_{max}(kW)}$}}&\multicolumn{1}{c|}{\textbf{$\mathbf{n_{init}}$}}&\multicolumn{1}{c|}{\textbf{$\mathbf{n_{min}}$}}&\multicolumn{1}{c|}{\textbf{$\mathbf{n_{max}}$}}\\\hline
        \multicolumn{1}{|c|}{\textbf{$\mathbf{DER_{18}}$} } &210&300 & \multicolumn{1}{|c|}{\multirow{4}{*}{0}} & \multicolumn{1}{|c|}{\multirow{4}{*}{$-36$}} & \multicolumn{1}{|c|}{\multirow{4}{*}{$+36$}}\\ \cline{1-3}
        \multicolumn{1}{|c|}{\textbf{$\mathbf{DER_{23}}$}}  &100 &600 &\multicolumn{1}{|c|}{} &\multicolumn{1}{|c|}{}&\multicolumn{1}{|c|}{}\\ \cline{1-3}
        \multicolumn{1}{|c|}{\textbf{$\mathbf{DER_{24}}$}}  &250&600&\multicolumn{1}{|c|}{}&\multicolumn{1}{|c|}{}&\multicolumn{1}{|c|}{}\\ \cline{1-3}
        \multicolumn{1}{|c|}{\textbf{$\mathbf{DER_{33}}$}}  &150&300 &\multicolumn{1}{|c|}{}&\multicolumn{1}{|c|}{}&\multicolumn{1}{|c|}{}\\ \hline
\end{tabular}
\end{center}
\end{table}

The optimal control problem is formulated as a linear one taking advantage of the voltage sensitivity coefficients. The controlled variables are the bus node voltages and the control variables are the active and reactive power injections of the DER and the transformer's tap positions under the control of the DNO, $\mathbf{\Delta x}=[\mathbf{\Delta P}_{DER}, \mathbf{\Delta Q}_{DER},\mathbf{\Delta n}]$. It is important to state that, formally, this problem is a mixed integer optimization problem due to the tap positions of the transformers. However, for reasons of simplicity, the tap positions are considered pseudo-continuous variables which are rounded to the nearest integer once the optimal solution is reached. The objective of the linear optimization problem relevant to the problem is:
\begin{align}
\min_{\Delta{\mathbf{x}}}\parallel \bar{E}_i-\bar{E}\parallel
\end{align}	
The linearized relationship that links bus voltages with control variables is expressed in the following way (e.g. \cite{borghetti2010short}):
\begin{align}
\Delta|\bar{E}_i|=\mathbf{K_{P}}_{i} \mathbf{\Delta P}_{i}+\mathbf{K_{Q}}_{i} \mathbf{{\Delta Q}}_{i}+\mathbf{K_{n}}_{i} \mathbf{{\Delta n}}_{i}
\end{align}
where $\mathbf{K_{P}}_{i}$ is the vector of sensitivity coefficients with respect to the active powers of the DERs, $\mathbf{K_{Q}}_{i}$ is the vector of sensitivity coefficients with respect to the reactive powers of the DERs and $\mathbf{K_{n}}_{i}$ is the vector of sensitivity coefficients with respect to the transformer's tap positions.
The imposed constraints on the operational points of the DERs and the tap positions are the following:
\begin{align}
0&\leq P_{DER_{i}}\leq P_{DER_{i_{max}}} \\ \nonumber
Q_{DER_{i_{min}}}&\leq Q_{DER_{i}}\leq Q_{DER_{i_{max}}}\\ \nonumber
n_{min}&\leq n \leq n_{max}
\end{align}	
In order to simplify the analysis, we have assumed that the DER capability curves are rectangular ones in the PQ plane.

The formulated linearized problem is solved by using the linear least squares method. The method used to calculate analytically the sensitivity coefficients allows us to consider two different optimization scenarios. In the first ($opt_1$), the operator of the system is assumed to control the set points of the DERs considering that they are injecting equal powers into the three phases, whereas in the second case ($opt_2$) it is assumed to have a more sophisticated control on each of the phases independently except for the tap-changers positions. It is worth noting that this second option, although far from a realistic implementation, allows us to show the capability of the proposed method to deal with the inherent unbalanced nature of distribution networks. Table~\ref{table:OPT1} and Table~\ref{table:OPT2} show the optimal operational set points corresponding to these cases.
\begin{table}[H]
\begin{center}
\caption{Optimal operational set points of the DERs and the tap-changers in the 34 test node feeder when the system operator has control on their 3-phase output}
\label{table:OPT1}
\begin{tabular}{ c | c | c |c |}
        \cline{2-4}
        & \multicolumn{1}{c|}{\textbf{$\mathbf{P_{opt_{1}}}$(kW)}}&\multicolumn{1}{c|}{\textbf{$\mathbf{Q_{opt_1}}$(kVar)}} &\multicolumn{1}{c|}{\textbf{$\mathbf{n_{opt_{1}}}$}}\\ \hline
        \multicolumn{1}{|c|}{\textbf{$\mathbf{DER_{18}}$} }  &300 & 300 &\multicolumn{1}{|c|}{\multirow{4}{*}{-2}}\\ \cline{1-3}
        \multicolumn{1}{|c|}{\textbf{$\mathbf{DER_{23}}$}}   &600 & 600 &\multicolumn{1}{|c|}{}\\ \cline{1-3}
        \multicolumn{1}{|c|}{\textbf{$\mathbf{DER_{24}}$}}   &600 & 264.06&\multicolumn{1}{|c|}{}\\ \cline{1-3}
        \multicolumn{1}{|c|}{\textbf{$\mathbf{DER_{33}}$}}   &300 & -14.46&\multicolumn{1}{|c|}{}\\ \hline
\end{tabular}
\end{center}
\end{table}

\begin{figure}
\begin{center}
\leavevmode
\subfigure[IEEE 34 node test feeder - Voltage profile of phase $a$ of the busses.]{\includegraphics[width=1\linewidth,clip=true,trim= 0 70 0 70]{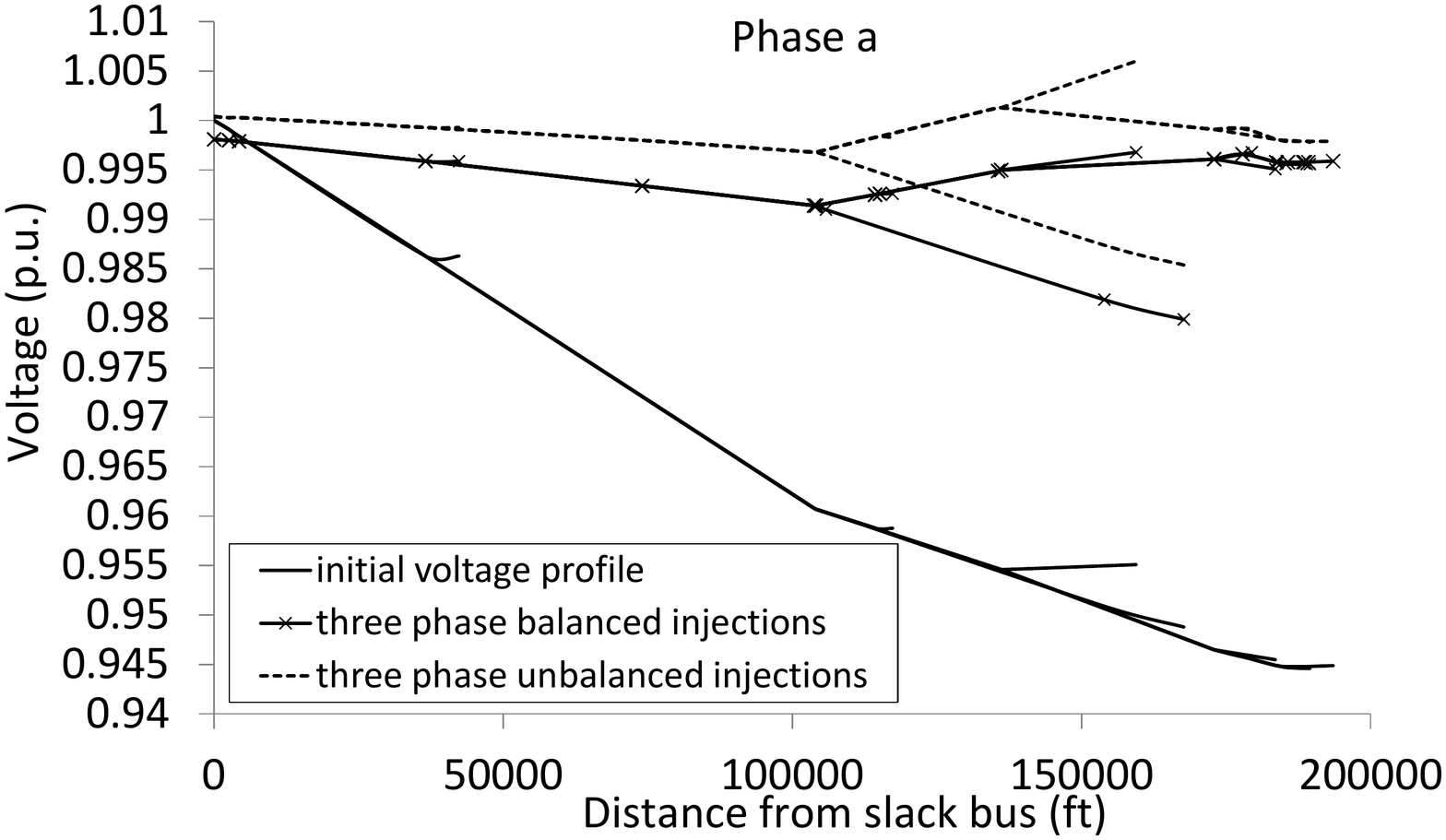}}
\subfigure[IEEE 34 node test feeder - Voltage profile of phase $b$ of the busses.]{\includegraphics[width=1\linewidth,clip=true,trim= 0 70 0 70]{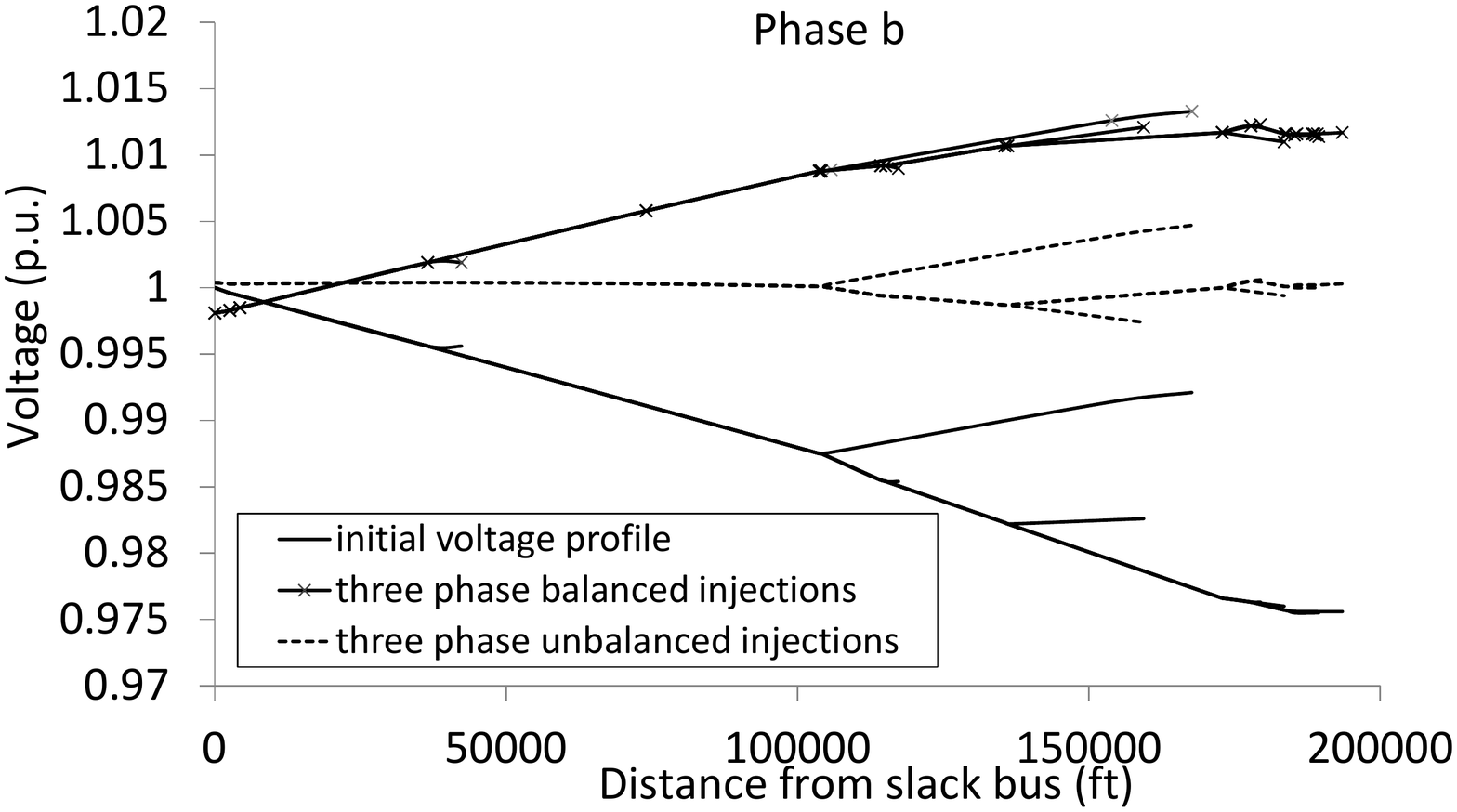}}
\subfigure[IEEE 34 node test feeder - Voltage profile of phase $c$ of the busses.]{\includegraphics[width=1\linewidth,clip=true,trim= 0 70 0 70]{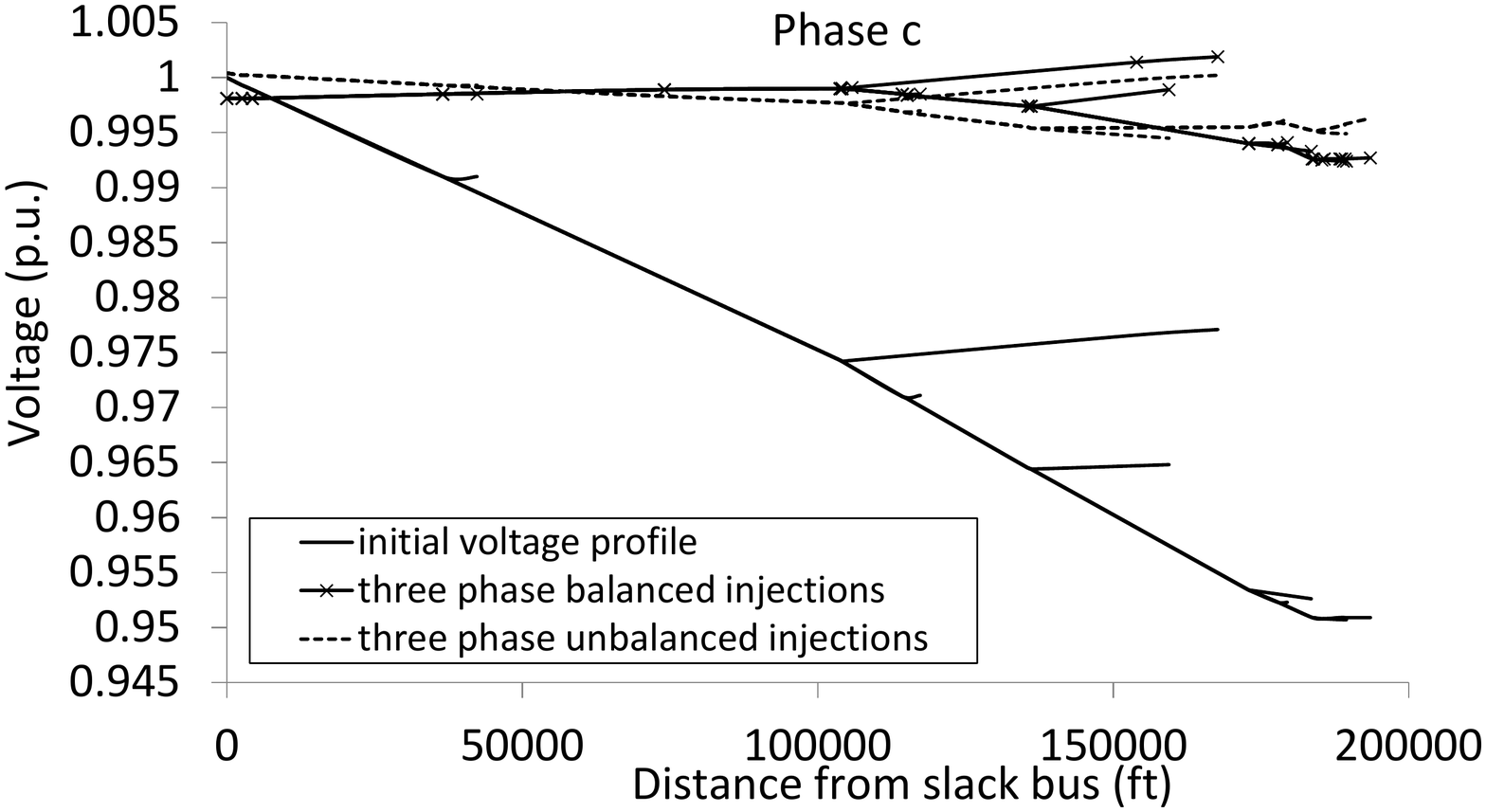}}
\end{center}
\caption{Initial and optimized voltage profile of the IEEE 34 node test feeder.} \label{fig:voltprof}
\end{figure}

Additionally, in Fig.\ref{fig:voltprof} the voltage profile of the busses of the system is presented in the initial and the optimal cases. The solid line in the figures shows the initial voltage profile, the solid line with the markers shows the first case optimal scenario ($opt_1$) and the dashed line represents the second case where the DNO has full control in each of the phases of the DERS ($opt_2$). The offset in the graphs, observed in the slack bus, depicts the optimal tap position in each case. What can be observed is that, when there is a possibility to control each of the three phases of the DERs output, the optimal voltage profile is better than the one corresponding to control of the 3-phase output of the set points of the DERs.

\begin{table}[H]
\begin{center}
\caption{Optimal operational set points of the DERs and the tap-changers in the 34 test node feeder when the system operator has control on each of the three phases independently}
\label{table:OPT2}
\begin{tabular}{ c | c | c |c |}
        \cline{2-4}
        & \multicolumn{1}{c|}{\textbf{$\mathbf{P{_opt_{2}}}$(kW)}}&\multicolumn{1}{c|}{\textbf{$\mathbf{Q_{opt_2}}$(kVar)}}&\multicolumn{1}{c|}{\textbf{$\mathbf{n_{opt_2}}$}}\\ \hline
        \multicolumn{1}{|c|}{\textbf{$\mathbf{DER^a_{18}}$}}  &100 & 100 &\multicolumn{1}{c|}{\multirow{12}{*}{+1}}\\ \cline{1-3}
        \multicolumn{1}{|c|}{\textbf{$\mathbf{DER^b_{18}}$}}  &100 & -88.56 &\multicolumn{1}{c|}{}\\ \cline{1-3}
        \multicolumn{1}{|c|}{\textbf{$\mathbf{DER^c_{18}}$}}  &0 & 83 &\multicolumn{1}{c|}{}\\ \cline{1-3}
        \multicolumn{1}{|c|}{\textbf{$\mathbf{DER^a_{23}}$}}   &200 &200  &\multicolumn{1}{c|}{}\\ \cline{1-3}
        \multicolumn{1}{|c|}{\textbf{$\mathbf{DER^b_{23}}$}}   &200 & 200 &\multicolumn{1}{c|}{}\\ \cline{1-3}
        \multicolumn{1}{|c|}{\textbf{$\mathbf{DER^c_{23}}$}}   &0 & 200 &\multicolumn{1}{c|}{}\\ \cline{1-3}
        \multicolumn{1}{|c|}{\textbf{$\mathbf{DER^a_{24}}$}}   &200 & 102.81 &\multicolumn{1}{c|}{}\\ \cline{1-3}
        \multicolumn{1}{|c|}{\textbf{$\mathbf{DER^b_{24}}$}}   &196.51 & 200 &\multicolumn{1}{c|}{}\\ \cline{1-3}
        \multicolumn{1}{|c|}{\textbf{$\mathbf{DER^c_{24}}$}}   &111.40& 200 &\multicolumn{1}{c|}{}\\ \cline{1-3}
        \multicolumn{1}{|c|}{\textbf{$\mathbf{DER^a_{33}}$}}   &100 & -27.88 &\multicolumn{1}{c|}{}\\ \cline{1-3}
        \multicolumn{1}{|c|}{\textbf{$\mathbf{DER^b_{33}}$}}   &100 & 100 &\multicolumn{1}{c|}{}\\ \cline{1-3}
        \multicolumn{1}{|c|}{\textbf{$\mathbf{DER^c_{33}}$}}   &98.40 & 100 &\multicolumn{1}{c|}{}\\\hline
\end{tabular}
\end{center}
\end{table}

%% file: 06_conclusion.tex
\section{Conclusion}
\label{sec-conclusion}
In this paper we have proposed a new method for the analytical computation of voltages and currents sensitivity coefficients as a function of the nodal power injections. The contributions of the proposed method are the following: (i) it is generalized to account for a generic number of slack busses; (ii) it allows the computation of sensitivities w.r.t. tap-changer positions (iii) it is proved to admit a unique solution for the case of radial networks and (iv) it supports the computation of the sensitivities for a generic unbalanced electrical network by using the $[\mathbf{Y}]$ compound matrix being, thus, suitable for distribution systems.

Compared to the traditional use of the Jacobian load-flow matrix, it allows us to reduce the computation time by almost a factor of three, thus enabling, in principle, its implementation in real-time optimal controllers.

The paper has also validated the proposed method by making reference to typical IEEE 13 and 34 nodes distribution test feeders. The former has been used to numerically validate the computation of the coefficients whilst the latter has been used to show an application example related to a possible integration of the proposed method for the problem of optimal voltage control in unbalanced distribution systems.

It is worth observing that the proposed analytical computation of voltages and currents sensitivities enables the reduction of the computational time of several traditional power systems problems involving non-negligible computational efforts, such as real-time centralized controls, contingency analysis or optimal planning.